\documentclass[10pt,twocolumn,twoside]{IEEEtran}

\usepackage{graphicx}
\usepackage{subfigure}
\usepackage{color}
\usepackage{multicol}
\usepackage{amssymb}
\usepackage{multirow}
\usepackage{amsmath}
\usepackage{bm}
\usepackage{cite}
\usepackage{gensymb}
\usepackage{amsfonts}
\usepackage{mathrsfs}
\usepackage{amsmath}
\usepackage{algorithm}
\usepackage{algorithmic}
\usepackage{amsthm}

\usepackage{tabularx}
\usepackage{balance}
\usepackage{mathrsfs}
\usepackage{array}
\usepackage{multirow}
\newcommand{\PreserveBackslash}[1]{\let\temp=\\#1\let\\=\temp}
\newcolumntype{C}[1]{>{\PreserveBackslash\centering}p{#1}}
\newcolumntype{L}[1]{>{\PreserveBackslash\raggedright}p{#1}}
\newcolumntype{R}[1]{>{\PreserveBackslash\raggedleft}p{#1}}
\usepackage[flushleft]{threeparttable}
\newcommand{\mb}[1]{{  \mathbf  #1}}  

\newtheorem{thm}{Lemma}

\begin{document}
	
	\bibliographystyle{IEEEtran} 
	
	\title{Channel Estimation for Extremely Large-Scale MIMO: Far-Field or Near-Field?}
	\author{\IEEEauthorblockN{ Mingyao Cui, \emph{Student Member, IEEE}, Linglong Dai, \emph{Fellow, IEEE}}

	\thanks{
		Part of this work has been accepted by IEEE Global Communications Conference (IEEE GLOBECOM'21) \cite{Cui2112:Near}.
		
		All authors are with the Beijing National Research Center for Information Science and Technology (BNRist) as well as the Department of Electronic Engineering, Tsinghua University, Beijing 100084, China (e-mails: cmy20@mails.tsinghua.edu.cn, daill@tsinghua.edu.cn).
	
This work was supported in part by the National Key Research and Development Program of China (Grant No.2020YFB1807201), in part by the National Natural Science Foundation of China (Grant No. 62031019), and in part by the European Commission through the H2020-MSCA-ITN META WIRELESS Research Project under Grant 956256. \emph{(Corresponding author: Linglong Dai.)}
}
\vspace{-4mm}
}
	\maketitle
	\IEEEpeerreviewmaketitle
	\begin{abstract}
Extremely large-scale multiple-input-multiple-output (XL-MIMO) is promising to meet the high rate requirements for future 6G. To realize efficient precoding, accurate channel state information is essential. Existing channel estimation algorithms with low pilot overhead heavily rely on the channel sparsity in the angular domain, which is achieved by the classical far-field planar-wavefront assumption. However, due to the non-negligible near-field spherical-wavefront property in XL-MIMO, this channel sparsity in the angular domain is not achievable. Therefore, existing far-field channel estimation schemes will suffer from severe performance loss. To address this problem, in this paper, we study the near-field channel estimation by exploiting the polar-domain sparsity. Specifically, unlike the classical angular-domain representation that only considers the angular information, we propose a polar-domain representation, which simultaneously accounts for both the angular and distance information. In this way, the near-field channel also exhibits sparsity in the polar domain, based on which, we propose on-grid and off-grid near-field XL-MIMO channel estimation schemes. Firstly, an on-grid polar-domain simultaneous orthogonal matching pursuit (P-SOMP) algorithm is proposed to efficiently estimate the near-field channel. Furthermore, an off-grid polar-domain simultaneous iterative gridless weighted (P-SIGW) algorithm is proposed to improve the estimation accuracy. Finally, simulations are provided to verify the effectiveness of our schemes.
	\end{abstract}
	
	\begin{IEEEkeywords}
		Near-field, XL-MIMO, hybrid precoding, channel representation, channel estimation.\vspace{-3mm}
	\end{IEEEkeywords}	
	\section{Introduction}
	
	Massive multiple-input-multiple-output (MIMO) is one of the most critical technologies for current 5G communications  \cite{5Gwork_Par13}.
	Equipped with massive antenna arrays at the base station (BS),  massive MIMO can improve the spectral efficiency by orders of magnitude through beamforming or multiplexing. 
	For future 6G communications, extremely large-scale MIMO (XL-MIMO), where the number of antennas can be much larger than that for massive MIMO  \cite{MIMO2_Hu2018}, can effectively achieve 10-fold increases in spectral efficiency \cite{6Gchallenge_Rappaport2019}.
	On the other hand, benefiting from the rich spectrum resource at  millimeter-wave (mmWave) band or terahertz (THz) band, high-frequency communications can provide largely available bandwidth \cite{THzsurvey_Elayan2020}.
	Meanwhile, the very small size of  high-frequency antennas favorably enables the deployment of XL-MIMO with an extremely large number of antennas. 
	Therefore, high-frequency XL-MIMO has been widely considered as a key enabling technology for future 6G \cite{6Gchallenge_Rappaport2019}.
	
	Similar to the current 5G mmWave massive MIMO, hybrid precoding architecture is widely considered for high-frequency XL-MIMO  \cite{NomaSWIPT_Dai19}, 	since the power consumption of the high-frequency radio-frequency (RF) chain is very high \cite{UMIMO_Han21}.
 	Efficient hybrid precoding requires accurate channel state information at the base station. 
 	Unfortunately, since the number of RF chains in hybrid precoding is much smaller than the number of antennas, the BS cannot observe the signals at each antenna simultaneously \cite{SWOMP_Robert18}. 
	This will result in unacceptable pilot overhead, especially when the number of antennas is huge in XL-MIMO systems.
	
	\subsection{Prior works}
	
	To alleviate the high pilot overhead in the channel estimation, in current 5G massive MIMO systems, by exploiting the channel sparsity in the angular domain, some compressive sensing (CS) based algorithms have been studied to accurately estimate the high-dimensional channels with low pilot overhead \cite{CE_OMP_Lee16, CE_SOMP_Gao16, SWOMP_Robert18, CE_SMP_Huang19, LAMP_Wei21, WidebeamCE_Gao2019}. 	  
	For example, by utilizing the angular-domain sparsity,  the classical orthogonal matching pursuit (OMP) algorithm was used in \cite{CE_OMP_Lee16} to recover the angular-domain channel, where the channel was transformed into its angular-domain representation through the standard spatial Fourier transform.
	As for wideband systems, a simultaneous OMP (SOMP) algorithm was proposed in \cite{CE_SOMP_Gao16}, which jointly recovered the channels at different subcarriers based on the common support assumption, i.e., the sparse supports in the angular domain at different subcarriers were assumed to be the same. 
	Moreover, by taking into account the colored noise induced by hybrid precoding, the pre-whitening procedure was introduced  in the SOMP algorithm \cite{SWOMP_Robert18}. Furthermore, the message passing (MP) algorithms are utilized in \cite{CE_SMP_Huang19, LAMP_Wei21} to recover sparse angular-domain channel with prior information.
	Note that all solutions above assumed that the angle of departures/arrivals (AoDs/AoAs) lie in discrete points in the angular domain (i.e., “on-grid” AoDs/AoAs), while the actual AoDs are continuously distributed (i.e., “off-grid” AoDs/AoAs) in practical systems. 
	To solve the resolution limitation of these on-grid algorithms, several off-grid solutions were studied in \cite{UGSWOMP_Ro17,TD_Zhou17, CESR_Hu18,WideCT_Go21}. For instance, \cite{WideCT_Go21} proposed a  simultaneous iterative gridless weighted (SIGW) algorithm to directly estimate the channel parameters, including the angles and the path gains, and thus the channel estimation accuracy could be improved.
 	
 	It is worth noting that all solutions above heavily rely on the channel sparsity in the angular domain. However, this channel sparsity may not be achievable in XL-MIMO systems, and thus existing channel estimation schemes cannot be directly applied to XL-MIMO.
 	Specifically, the change from massive MIMO to XL-MIMO  not only means the increase in antenna
 	number, but also leads to the fundamental change in the electromagnetic field structure.
 	The radiation field of the electromagnetic wave can be divided into two regions, i.e., the far-field region and the near-field region \cite{fresnel_Selvan2017}.
 	The boundary to divide the near-field and the far-field is determined by the Rayleigh distance \cite{fresnel_Selvan2017},  which is  proportional to the square of the number of antennas. 
 	In current 5G massive MIMO systems, as the number of antennas is not very large, the Rayleigh distance is usually several meters, which is negligible in practice. Therefore, the wavefront can be simply modeled under the far-field \emph{planar wavefront} assumption, which only depends on the angle of departure/arrival (AoD/AoA) of the channel. Note that the channel sparsity in the angular domain is derived from this planar wavefront assumption.
 	In future 6G XL-MIMO systems, due to the significant increase in the number of antennas, the Rayleigh distance can be up to several hundreds of meters, thus the near-field region in XL-MIMO systems becomes no longer negligible. 
 	When the receiver is located in the near-field region, the wavefront should be accurately modeled under the \emph{spherical wavefront} assumption \cite{NearLoS_Zhou2015}, which is decided by not only the AoD/AoA but also the distance between the BS and the scatter.
 	For this near-field channel, a severe energy spread effect will be introduced for the classical angular-domain channel representation, i.e., the energy of a single near-field path component will be spread into multiple angles. 
 	In this case, the angular-domain channel may not be sparse, and thus existing far-field channel estimation  schemes based on the channel sparsity in the angular domain will suffer from severe performance degradation. 
	Consequently, to support the ultra-high data rate for future 6G, the efficient near-field channel estimation algorithm is essential for practical high-frequency XL-MIMO systems.


	Unfortunately, up to now, there are no related works on  near-field channel estimation for high-frequency XL-MIMO. Under the near-field spherical wavefront conditions, there are some existing works \cite{SPlocalization_Wei18,  NearCE_Liu2020,NearCE_Fried2019,NearCE_Han2020} investigating a similar problem in wireless sensing systems, i.e., the near-field localization problem. For instance, by exploiting the structure of the signal covariance matrix, several high-order statistic based methods, (e.g., the subspace-based algorithms \cite{SPlocalization_Wei18, NearCE_Liu2020} or high-order MUSIC algorithms \cite{NearCE_Fried2019}) have been proposed to estimate the AoD and distance between the source and the receiver in the near-field. 
	Furthermore, the OMP algorithm was improved in \cite{NearCE_Han2020} to locate near-field scatters. 
	However, all near-field localization methods above assume that the dimension of received signals is equal to or larger than that of the channel, which is not valid in the practical high-frequency XL-MIMO systems.
	To the best of our knowledge, the near-field channel estimation in high-frequency XL-MIMO systems has not been studied in the literature.

	\subsection{Our contributions}
	To fill in this gap, in this paper, the important problem of near-field channel estimation for XL-MIMO is studied, which is realized by replacing the classical angular-domain representation with a polar-domain representation\footnote{Simulation codes are provided to reproduce the results in this paper: http://oa.ee.tsinghua.edu.cn/dailinglong/publications/publications.html.}. Specifically, our contributions are summarized as follows.
	\begin{itemize}
		\item Firstly, by comparing the difference between the far-field and near-field channels, we reveal the energy spread effect when the practical near-field channel is transformed into the angular domain by using the classical spatial Fourier matrix. This energy spread effect indicates that the energy of a single near-field path component will be spread into multiple angles, and thus the near-field channel in the angular domain is no longer sparse.
		\item To deal with the energy spread effect, we propose a polar-domain representation of the near-field channel. Unlike the angular-domain representation of the far-field channel that only considers the angular information, the polar-domain representation accounts for both the angular and distance information simultaneously.  
		To design the polar-domain transform matrix, we utilize the Fresnel approximation to approximate the near-field channel. Then, based on the Fresnel function, we analyze how to sample the angle and distance in the polar domain to reduce the column coherence of the transform matrix. 
		Analytical results show that, the angle should be sampled uniformly, while the distance should be sampled non-uniformly. 
		We further point out that both the far-field and near-field channels exhibit sparsity in the polar domain, and thus the classical angular-domain representation is a special case of the proposed polar-domain representation.
		\item By exploiting the channel sparsity in the polar domain, an on-grid polar-domain simultaneous orthogonal matching pursuit (P-SOMP) algorithm is proposed to estimate the near-field channel efficiently. To further improve the estimation accuracy, we propose an off-grid polar-domain simultaneous iterative gridless weighted (P-SIGW) algorithm, where the near-field channel parameters are directly estimated. Unlike existing off-grid channel estimation algorithms \cite{UGSWOMP_Ro17, TD_Zhou17, CESR_Hu18, WideCT_Go21} that only estimate path gains and angles, the proposed P-SIGW algorithm simultaneously recovers the path gains, angles, and distances. Finally, numerical results are provided to verify the effectiveness of the proposed algorithms.
	\end{itemize}  
	
	\subsection{Organization and notation}
	\emph{Organization}: The remainder of this paper is organized as follows. In section \ref{sec:2}, the system model is introduced, and the energy spread effect is revealed. In section \ref{sec:3}, the polar-domain representation is proposed, and the method to design the polar-domain transform  matrix is provided. 
	In section \ref{sec:4}, the on-grid P-SOMP algorithm and the off-grid P-SIGW algorithm are proposed. Simulations are carried out in Section \ref{sec:5}, and finally conclusions are drawn in Section \ref{sec:6}.
	
	\emph{Notation}: Lower-case and upper-case boldface letters represent vectors and matrices, respectively; $X_{p,q}$ denotes the $(p, q)$-th entry of the matrix $\mb{X}$; $\mb{X}_{p, :}$ and $\mb{X}_{:, p}$ denote the $p$-th row and the $p$-th column of the matrix $\mb{X}$; $(\cdot)^T$ and $(\cdot)^H$ denote the transpose and conjugate transpose, respectively; $|\cdot|$ denotes the absolute
	operator;  $\text{Tr}(\cdot)$ denotes the trace operator; $\mathcal{CN}(\mu, \Sigma)$ and $\mathcal{U}(a,b)$
	denote the Gaussian distribution with mean $\mu$ and covariance
	$\Sigma$, and the uniform distribution between $a$ and $b$, respectively. 
	\section{System Model} \label{sec:2}

\begin{figure}
	\centering
	\includegraphics[width=3.5in]{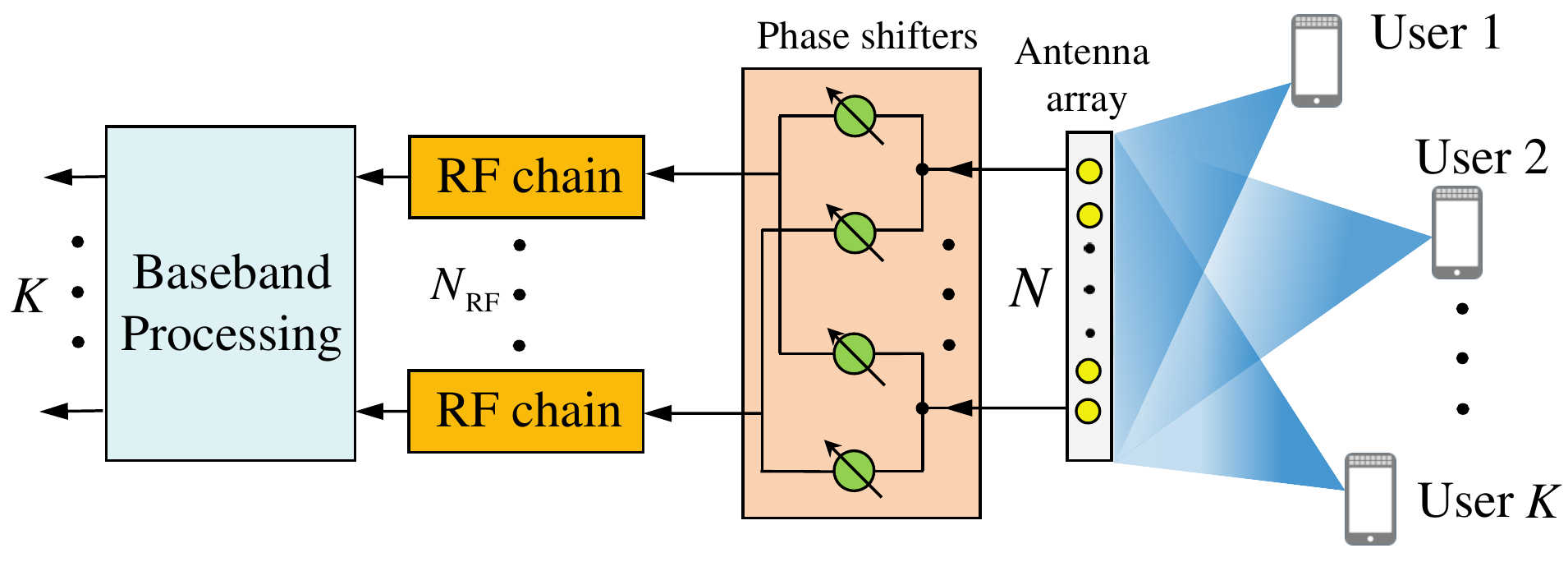}
	\vspace*{-1em}
	\caption{ XL-MIMO system with hybrid precoding.
	}
	\label{img:layout}
	\vspace*{-1em}
\end{figure}

	As shown in Fig. \ref{img:layout}, we consider an uplink time division duplexing (TDD) based  XL-MIMO OFDM communication system in this paper. The hybrid precoding architecture is employed at the BS. {The BS is equipped with $N_{\text{RF}}$ RF chains and an $N$-antenna uniform linear array, where $N_{\text{RF}} \ll N $.} The antenna spacing is $d = \frac{\lambda_c}{2}$, where $\lambda_c$ is the carrier wavelength. $K$ single-antenna users are served with $M$ subcarriers simultaneously, where $K \le N_{\text{RF}}$. 
	{ For uplink channel estimation, we assume the $K$ users transmit mutual orthogonal pilot sequences to the BS \cite{FundWC_Tse2015}, e.g., orthogonal time or frequency resources are utilized for different users to transmit pilot sequences. Therefore, channel estimation for each user is independent. Without loss of generality, we consider an arbitrary user. }

	Specifically, we denote $x_{m,p}$ as the transmit pilot at the $m$-th subcarrier in time slot $p$. Then, the received pilot $\mb{y}_{m,p} \in \mathbb{C}^{N_{\text{RF}} \times 1}$ is 
	\begin{align}
	  \mb{y}_{m,p} = \mb{A}_p \mb{h}_m x_{m,p} + \mb{A}_p \mb{n}_{m,p},
	\end{align}
	where $\mb{A}_p \in \mathbb{C}^{N_{\text{RF}}\times N}$ denotes the analog combining matrix satisfying the constant modulus constraint $| \mb{A}_p(i,j)| = \frac{1}{\sqrt{N}}$, and $\mb{n}_{m,p} \in \mathbb{C}^{N \times 1}$ denotes the Gaussian complex noise following the distribution $\mathcal{CN}(0, \sigma^2\mb{I}_N)$. Define $P$ as the pilot length and assume $x_{m,p} = 1$ for $p = 1,2,\cdots,P$. Then the overall received pilot sequence $\mb{y}_m = \left[ \mb{y}^T_{m,1}, \cdots, \mb{y}^T_{m, P} \right]^T $ at the $m$-th subcarrier can be denoted as 
	\begin{align} \label{eq:HP}
	\mb{y}_m = \mb{A} \mb{h}_m + \mb{n}_m,
	\end{align}
	where $\mb{n}_m = \left[ \mb{n}^T_{m,1} \mb{A}^T_{1}, \cdots, \mb{n}^T_{m,P}\mb{A}^T_{P} \right]^T$ denotes the noise. {$\mb{A} = \left[ \mb{A}^T_{1}, \cdots, \mb{A}^T_{P} \right]^T \in \mathbb{C}^{PN_{\text{RF}} \times N}$ denotes the overall observation matrix, where the elements in $\mb{A}$ are independent and can be randomly generated from the set $\frac{1}{\sqrt{N}}\{-1,1\}$ with equal probability. }
	Since the BS antenna number $N$ is very large, the dimension $PN_{\text RF}$  of the received signal $\mb{y}_m$ is usually much lower than $N$, which makes it challenging to estimate $\mb{h}_m$ from $\mb{y}_m$. 
	
	Fortunately, the channel sparsity in the angular domain at high-frequency enables the compressive sensing (CS) based channel estimation methods, where the pilot length $P$ can be significantly reduced.
	Specifically, users are assumed to be in the far-field region of the BS, where the channel is modeled under planar-wave assumption. 
	At the $m$-th subcarrier, the classical far-field channel is expressed as \cite{CE_Wang21}
	\begin{align}\label{eq:farchannel}
	\mb{h}_m^{\text{far-field}} = \sqrt{\frac{N}{L}}\sum_{l=1}^{L}g_l e^{-jk_mr_l}\mb{a}(\theta_l),
	\end{align}
	where $k_m = \frac{2\pi f_m}{c}$ denotes the wavenumber, $L$ is the number of paths. Moreover, $g_l$, $r_l$, and $\theta_l$ are the complex path gain, the distance, and the angle of the $l$-th path, respectively.  The steering vector $\mb{a}(\theta_l)$ on the angle $\theta_l \in [-1, 1]$ is derived from the planar-wave assumption, which is expressed as 
	\begin{align}\label{eq:ffsv}
	\mb{a}(\theta_l) = \frac{1}{\sqrt{N}}[1, e^{j\pi\theta_l},\cdots, e^{j(N-1)\pi\theta_l}]^T	.
	\end{align}

\begin{figure}
		\centering
		\includegraphics[width=3.5in]{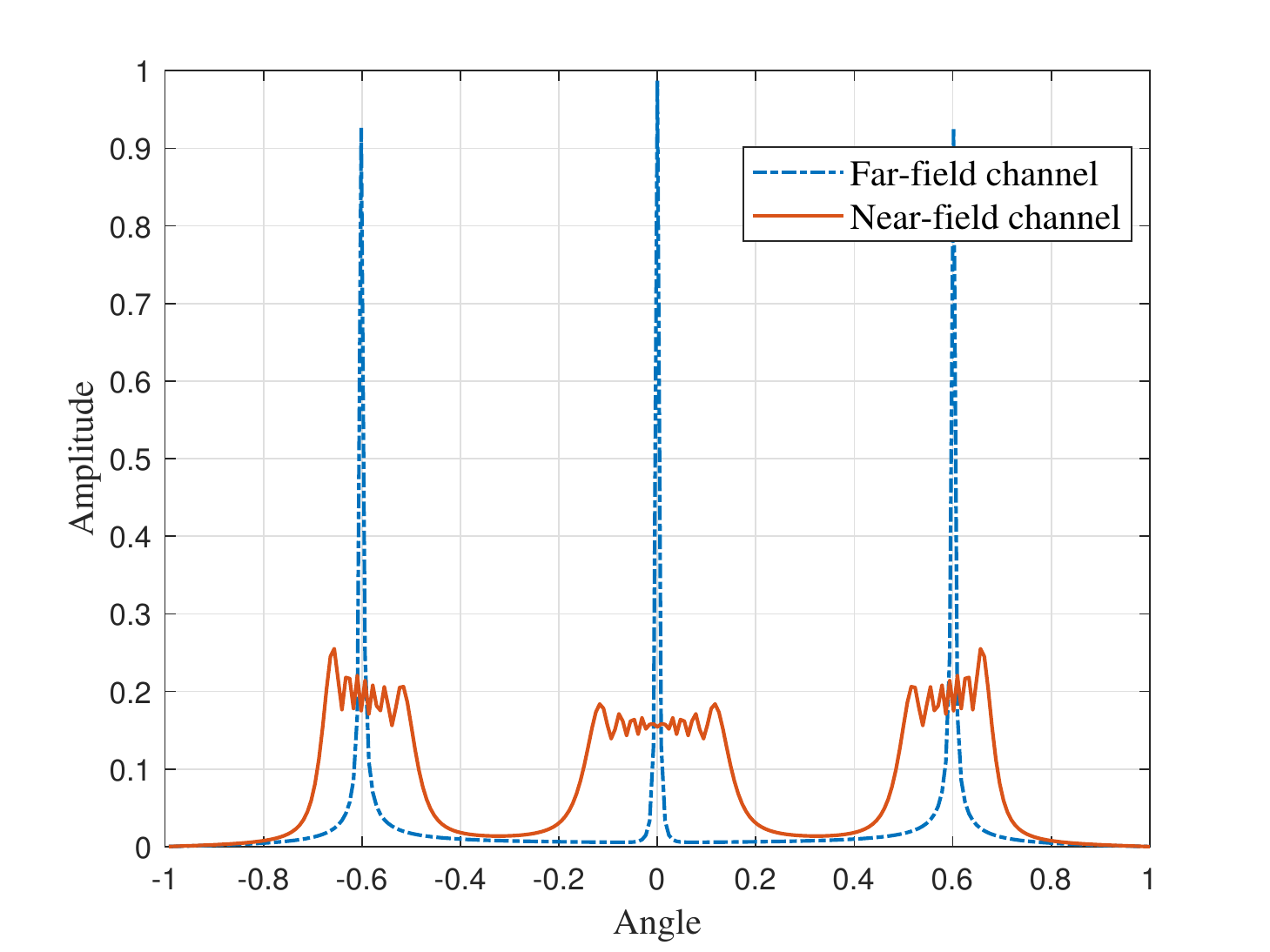}
		\vspace*{-1em}
		\caption{ Comparison of the angular-domain representations between the far-field and near-field channels. The BS antenna number is $256$ and the carrier is $100$ GHz. For far-field, the user is 100 meters from the BS, and for near-field, the user is 5 meters away from the BS while $L = 3$ paths are chosen.
		}
		\label{img:angle domain}
		\vspace*{-1em}
	\end{figure}

	Note that the phase of each element in the steering vector $\mb{a}(\cdot)$ is linear to the antenna index $n$, thus $\mb{a}(\cdot)$ is a discrete Fourier vector. 
	Correspondingly, the channel $\mb{h}_m^{\text{far-field}}$ can be transformed into its angular-domain representation $\mb{h}^{\mathcal{A}}_m$ by the spatial  Fourier  transform matrix \cite{CE_SMP_Huang19}, i.e., 
	\begin{align}\label{eq:at}
		\mb{h}_m^{\text{far-field}} = \mb{F}\mb{h}^{\mathcal{A}}_m,
	\end{align}
	where $\mb{F} \in \mathbb{C}^{N\times N}$ denotes the Fourier transform matrix. $\mb{F}$ contains $N$ orthogonal steering vectors uniformly sampled from the whole angular space as $\mb{F} = [\mb{a}(\theta_0),  \cdots, \mb{a}(\theta_{N - 1})]$, where $\theta_n = \frac{2n-N+1}{N}$, $n=0,1,\cdots,N-1$. As shown in Fig. \ref{img:angle domain}, since the number of scatters is limited, the far-field channel $\mb{h}^{\mathcal{A}}_m$ in the angular domain is usually sparse. Utilizing this channel sparsity, some compressive sensing (CS) based channel estimation algorithms have been proposed to efficiently recover $\mb{h}_m^{\mathcal{A}}$  with low pilot overhead $P$  \cite{CE_OMP_Lee16, CE_SOMP_Gao16, SWOMP_Robert18, CE_SMP_Huang19, LAMP_Wei21, WidebeamCE_Gao2019}.  

\begin{figure}
		\centering
		\includegraphics[width=3.5in]{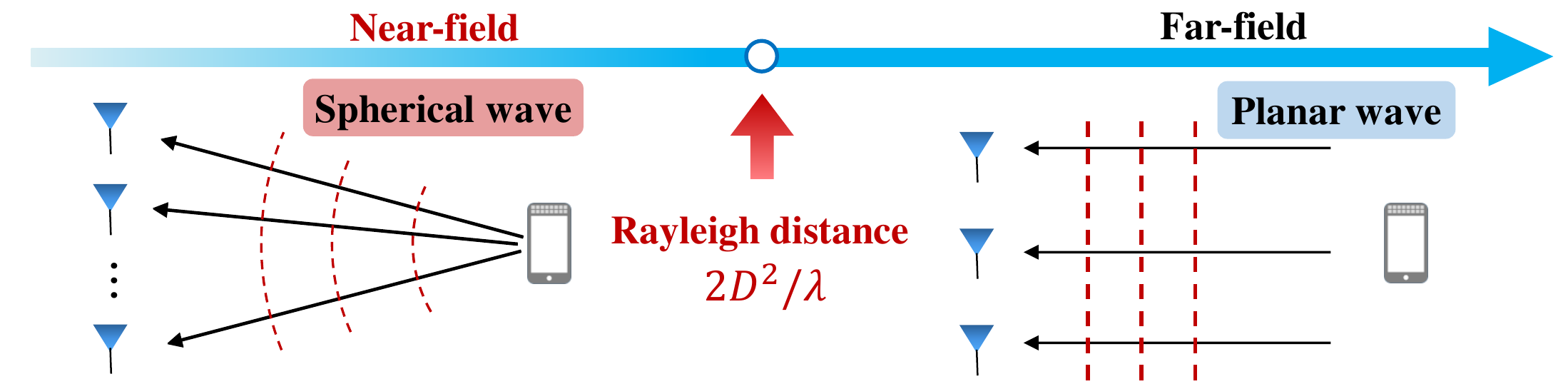}
		\caption{ The near-field region and the far-field region separated by the Rayleigh distance.
		}
		\label{img:RD}
		\vspace*{-1em}
	\end{figure}

	However, this channel sparsity in the angular domain may no longer be achievable in  XL-MIMO systems. The change from massive MIMO to XL-MIMO not only means the increase in antenna number, but also leads to the fundamental transformation of electromagnetic field structure. As shown in Fig. \ref{img:RD}, the radiation field of electromagnetic can be divided into two regions, i.e., the far-field and near-field regions.
{ 
	The widely adopted boundary between these two fields is  the Rayleigh distance  $Z = \frac{2D^2}{\lambda_c}$ \cite{fresnel_Selvan2017}, where $D$ and $\lambda_c$ denote the array aperture and wavelength, respectively. As the antenna spacing is $d = \frac{\lambda_c}{2}$ and the number of antennas is $N$, the array aperture of a uniform linear array is $D = Nd = N\frac{\lambda_c}{2}$. Therefore, the Rayleigh distance is $Z = \frac{1}{2}N^2\lambda_c$, which is proportional to $N^2$.
} 
The physical meaning of Rayleigh distance $Z$ is that, when the distance between the source and receiver is larger than $Z$,  the radiation field is far-field, and the wavefronts can be approximated as planar waves. Otherwise, if the distance between the radiating source and the receiver is less than $Z$, the radiation field is near-field, and the wavefronts are spherical waves.
	In current 5G massive MIMO systems, as the array aperture is
	not very large, the Rayleigh distance is usually several meters, which is negligible in practice. 
	However, in future 6G XL-MIMO systems, due to the significant
	increase in the number of antennas, the Rayleigh distance can be up to several hundreds of meters, thus the near-field region in XL-MIMO becomes not negligible. For instance, if the array aperture is $0.4$ m and the carrier is 100 GHz, then the Rayleigh distance is around 107 meters, which covers a large part of a cell.

	\begin{figure}
	\centering 
	\includegraphics[width=3.5in]{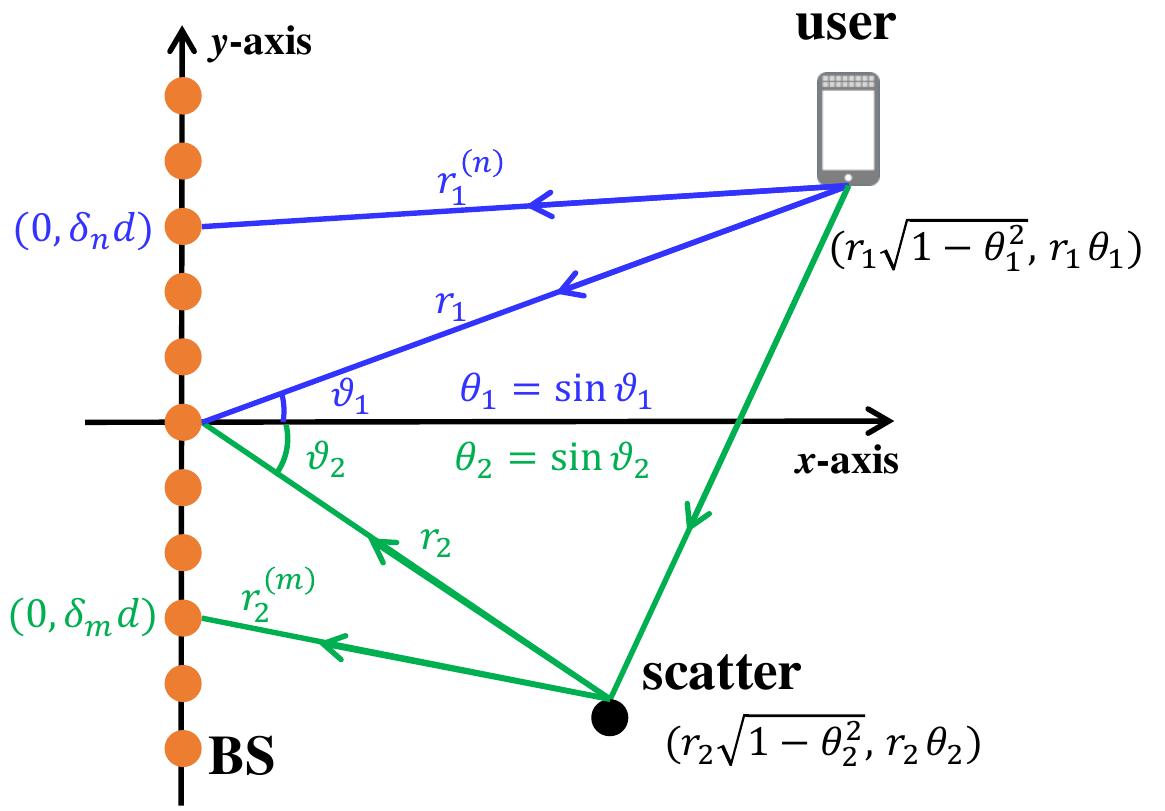}
	\caption{  The near-field channel model with two paths.
	} 
	\label{img:channel}
	\vspace*{-1em}
\end{figure}	
	The spherical wave channel model in near-field region can be presented as 	\cite{fresnel_Sherman1962}
	\begin{align}\label{eq:nearchannel}
	\mb{h}_m = \sqrt{\frac{N}{L}}\sum_{l=1}^{L}g_l e^{-jk_mr_l} \mb{b}(\theta_l, r_l).
	\end{align}
	The main difference between the far-field channel (\ref{eq:farchannel}) and the near-field channel (\ref{eq:nearchannel}) is the steering vector $\mb{b}(\cdot)$. The far-field steering vector $\mb{a}$ is derived from the planar-wave assumption, while the near-field steering vector $\mb{b}$ is derived from the accurate spherical wave, i.e.,  
{	\begin{align}\label{eq:nfsv}
	\mb{b}(\theta_l, r_l) = \frac{1}{\sqrt{N}}[e^{-jk_c (r_{l}^{(0)} - r_{l})},\cdots, e^{-jk_c (r_{l}^{(N-1)} - r_{l})}]^T,
	\end{align} 
	where $k_c = \frac{2\pi f_c}{c} = \frac{2\pi}{\lambda_c}$ denotes the wavenumber at the central carrier $f_c$. $r_l$ denotes the distance between the BS and the scatter or user, and $r_l^{(n)}$ denotes the distance between the $n$-th BS antenna and the scatter or user. The schematic diagram of near-field channel model is plotted in Fig. \ref{img:channel}. For expression simplicity, the channel is composed of a LOS path and an NLOS path. For the LOS path, $r_1^{(n)}$ denotes the distance between the $n$-th antenna and the user, while for the NLOS path, $r_2^{(m)}$ denotes the distance between the $m$-th antenna and the corresponding scatter.
	Suppose the coordinate of the $n$-th antenna is $(0, \delta_n d)$, where $\delta_n = \frac{2n - N + 1}{2}, n = 0,1,\cdots, N-1$, then it can be derived from the geometry that $r_{l}^{(n)} = \sqrt{ (r_l\sqrt{1 - \theta_1^2} - 0)^2 + (r_l\theta_l - \delta_nd)^2} = \sqrt{r_l^2 + \delta_n^2d^2 - 2r_l\theta_l\delta_nd}$, where $\theta_l \in [-1, 1]$ denotes the spatial angle.}
	This spherical wave model indicates that the phase of each element in the steering vector $\mb{b}(\cdot)$ is nonlinear to the antenna index $n$, so $\mb{b}(\cdot)$ is not a Fourier vector.
	In this case, $\mb{b}(\cdot)$ cannot be described by a single far-field Fourier vector.   As shown in Fig. \ref{img:angle domain}, several far-field Fourier vectors should be jointly utilized to describe a  near-field steering vectors $\mb{b}(\cdot)$. 
	Consequently, the energy of one near-field path component is no longer concentrated in one angle, but spread towards multiple angles, which is called the energy spread effect in this paper.
	This energy spread effect implies that in the near-field, the channel $\mb{h}_m^{\mathcal{A}}$ in the angular domain may not be sparse. Thus existing far-field channel estimation  schemes, based on angular-domain sparsity, will suffer from severe performance degradation in XL-MIMO systems.
		\section{Near-Field Polar-Domain Representation} \label{sec:3}
To realize efficient near-field channel estimation with reduced pilot overhead, in this section, we will propose a polar-domain representation of the XL-MIMO near-field channel to address the energy spread effect.
\subsection{Polar-domain representation for the near-field channel}
Although it is observed from Fig. \ref{img:angle domain} that the near-field channel is not sparse in the angular-domain, as shown in (\ref{eq:nearchannel}), the number of paths is still limited, i.e., $L \ll N$. This indicates that the number of channel parameters to be estimated is still limited, and thus the near-field channel is also compressible. 
	\begin{figure}
	\centering
	\subfigure[The angular-domain representation]
	{\includegraphics[width=3in]{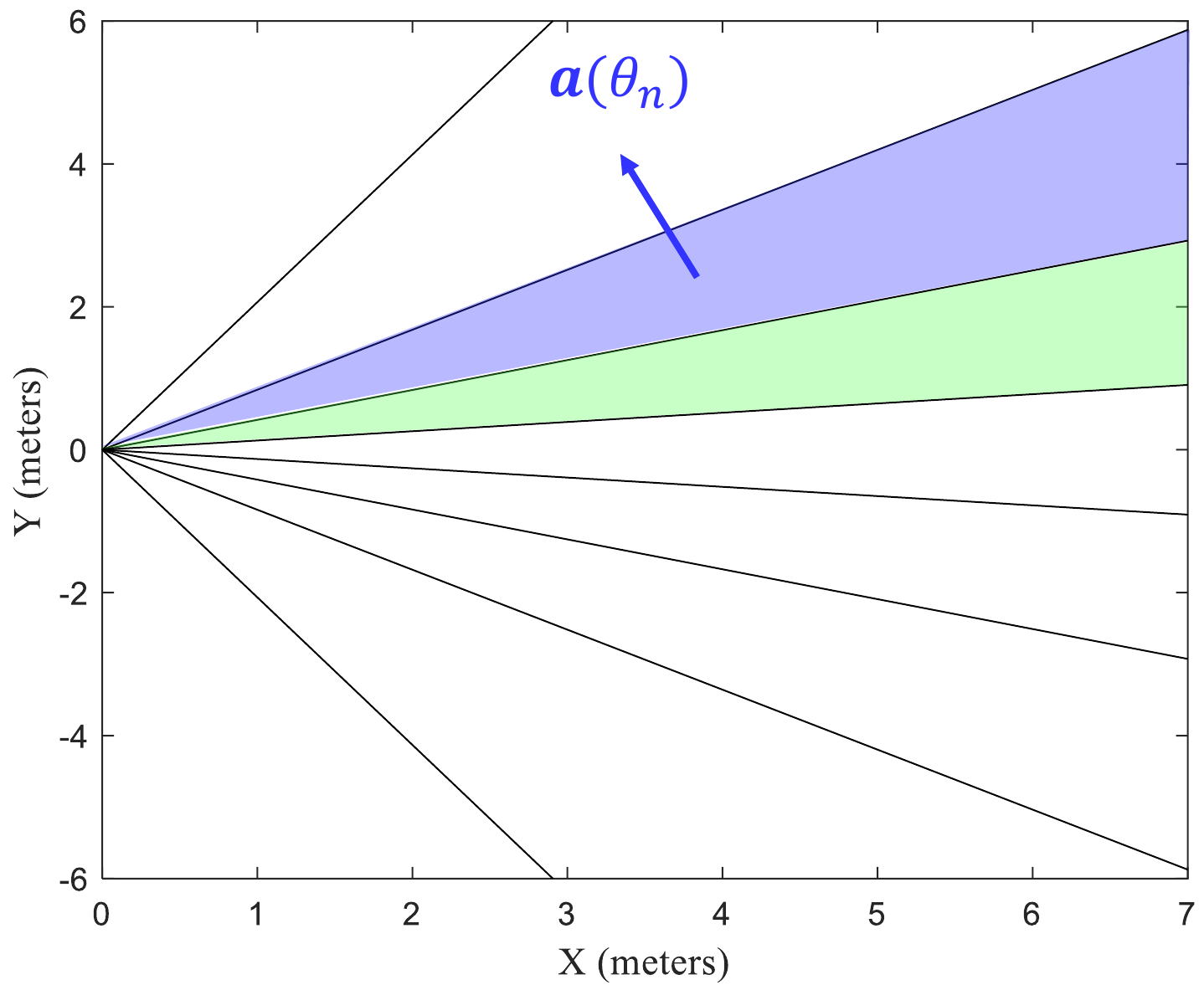}}
	\subfigure[The polar-domain representation]
	{\includegraphics[width=3in]{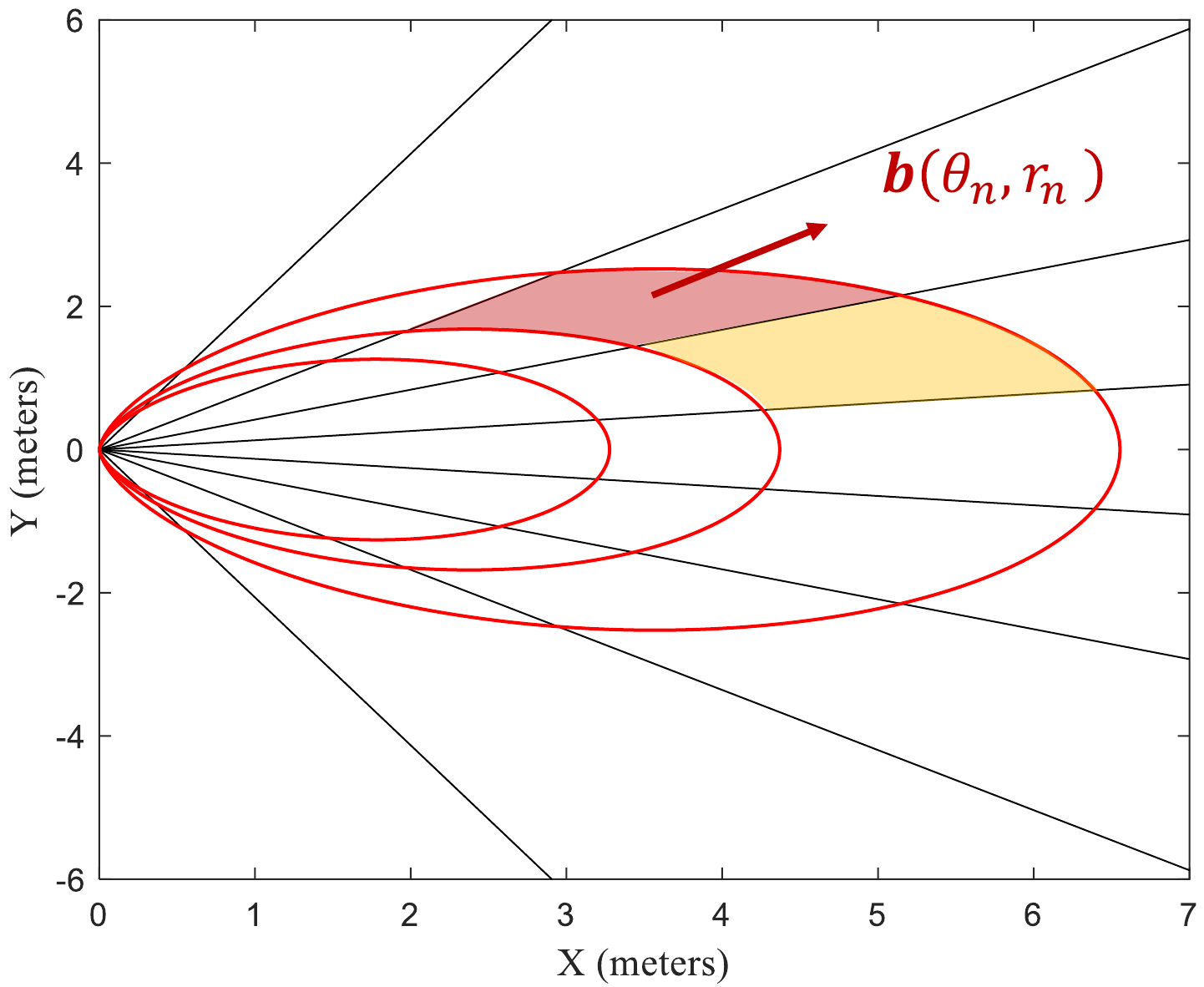}}
	\caption{Comparison between {(a)} the angular-domain representation and {(b)} the polar-domain representation.
	}
	\label{img:sampling}
	\vspace*{-1em}
\end{figure}

To find a sparse representation for the near-field channel, we can refer to the derivation of angular-domain sparse representation (\ref{eq:at}) for the far-field channel.
Specifically, as shown in (\ref{eq:farchannel}), the far-field channel $\mb{h}_m^{\text{far-field}}$ can be regarded as the weighted sum of  limited far-field steering vectors $\mb{a}(\theta)$, where $\mb{a}(\theta)$ only depends on the channel angle. Meanwhile,
the angular-domain transform matrix $\mb{F}$ in (\ref{eq:at}) is exactly composed of many far-field steering vectors $\mb{a}(\theta)$, where the angles $\theta$ are sampled from the entire angular domain to fully exploit the angular information of a far-field path component. Therefore, by utilizing this transform matrix $\mb{F}$, the angular-domain representation $\mb{h}_m^A$ is sparse in the far-field.

As for the sparse representation of the near-field channel $\mb{h}_m$, as shown in (\ref{eq:nearchannel}), $\mb{h}_m$ can be regarded as the weighted sum of limited near-field steering vectors $\mb{b}(\theta, r)$, where $\mb{b}(\theta, r)$ depends  not only the channel angle, but also the channel distance. 
Similar to the design of the existing matrix $\mb{F}$, we propose to design a new transform matrix $\mb{W}$, which is composed of many near-field steering vectors $\mb{b}(\theta, r)$, where the distances $r$ and angles $\theta$ are sampled from the entire angular-distance domain. 
In this way, the angular and distance information of a near-field path component is fully exploited by $\mb{W}$. 
Since angles and distances represent the coordinates in the polar coordinate system, the angular-distance domain is called the polar domain.
Accordingly, the matrix $\mb{W}$ is called as the polar-domain transform matrix. This idea is intuitively shown in Fig. \ref{img:sampling}.
Compared to the angular-domain transform matrix $\mb{F}$, which only samples different angles as shown in Fig. \ref{img:sampling} (a), the polar-domain transform matrix $\mb{W}$ simultaneously samples different angles and distances  as shown in Fig. \ref{img:sampling} (b).

{ Similar to the angular-domain representation $\mb{h}_m = \mb{F}\mb{h}_m^{\mathcal{A}}$, the proposed polar-domain representation $\mb{h}_m^{\mathcal{P}}$ of the near-field channel is 
\begin{align}\label{eq:pt}
\mb{h}_m = \mb{W}\mb{h}_m^{\mathcal{P}},
\end{align}
where $\mb{W} \in \mathbb{C}^{N \times Q}$, and $Q$ denotes the number of sampled near-field steering vectors in the polar domain.
Analog to the angular-domain sparsity where the angular information in far-field is considered by the matrix $\mb{F}$, the proposed polar-domain transform matrix $\mb{W}$ can account for both the angle and distance information of a near-field path component. Therefore, the near-field channel becomes sparse in the polar domain by utilizing the matrix $\mb{W}$. In this way, the energy spread effect for the near-field channel in the angular domain is avoided. Since the matrix $\mb{W}$ simultaneously samples different angles and distances, the number of sampled vectors $Q$ is not assumed to be equal to $N$, and generally, $Q$ is larger than $N$. Therefore, $\mb{W}$ is a wide matrix, but not a square matrix.
}

For the proposed polar-domain representation $\mb{h}_m = \mb{W}\mb{h}_m^{\mathcal{P}}$, a fundamental question is  how to sample the angles and distances to design the transform matrix $\mb{W}$? 
Following the CS framework\cite{CS_Ba10}, to achieve the satisfying channel recovery accuracy, the sampling on angle and distance should make the column coherence $\mu = \max_{p \neq q} \left|\mb{b}(\theta_p, r_p)^H\mb{b}(\theta_q, r_q)\right|$ of the polar-domain transform matrix $\mb{W}$ as small as possible, where $\mb{b}(\theta_p, r_p)$ and $\mb{b}(\theta_q, r_q)$ are two columns of $\mb{W}$.
However, since the phase $-k_c(r^{(n)} - r)$ of $n$-th element in $\mb{b}(\theta, r)$ is non-linear to the antenna index $n$ as shown in (\ref{eq:nfsv}), it is intractable to get a close form of $\left|\mb{b}(\theta_p, r_p)^H\mb{b}(\theta_q, r_q)\right|$, which makes the design of $\mb{W}$ difficult.


In the following discussions, we will approximately derive the column coherence between two near-field steering vectors, i.e., $f(\theta_p, \theta_q, r_p, r_q) = \left|\mb{b}(\theta_p, r_p)^H\mb{b}(\theta_q, r_q)\right|$.
Based on the Fresnel approximation \cite{fresnel_Sherman1962}, the distance  $r^{(n)}$ between the $n$-th antenna and the user or scatter can be approximated as $ 	r^{(n)} = \sqrt{r^2 - 2r\delta_nd\theta + \delta_n^2d} 
\overset{(a)}{\approx} r - \delta_nd\theta + \frac{\delta_n^2d^2(1 - \theta^2)}{2r}$, where (a) is derived by $\sqrt{1+x} \approx 1 + \frac{1}{2}x - \frac{1}{8}x^2$. 
{
	It has been studied in \cite{fresnel_Selvan2017} that the Fresnel approximation (a) is accurate when the distance between the BS and the user or scatter  is larger than $0.5 \sqrt{\frac{D^3}{\lambda_c}}$, which is much lower than the Rayleigh distance $\frac{2D^2}{\lambda_c}$. For instance, if the array aperture $D$ is $0.4$ m and the wavelength $\lambda_c$ is $3$ mm, then the Rayleigh distance is 106.7 m, but $0.5 \sqrt{\frac{D^3}{\lambda_c}}$ is only 2.3 m, which is nearly negligible.
}
Therefore, the column coherence $f(\theta_p, \theta_q, r_p, r_q)$ can be approximated as 
\begin{align} 
	&f(\theta_p, \theta_q, r_p, r_q) = \left|\frac{1}{N}\sum_{\delta_n}e^{jk_c(r_p(n) - r_q(n))}\right| \notag \\
	&\approx  \left|\frac{1}{N}\sum_{\delta_n}e^{jk_c\delta_nd (\theta_q - \theta_p) + jk_c\delta_n^2d^2\left( \frac{1-\theta_p^2}{2r_p} - \frac{1-\theta_q^2}{2r_q}\right) }\right| \notag\\
&	= \left|\frac{1}{N}\sum_{n=-(N-1)/2}^{(N-1)/2}e^{jn\pi (\theta_q - \theta_p) + jk_cn^2d^2\left( \frac{1-\theta_p^2}{2r_p} - \frac{1-\theta_q^2}{2r_q}\right) }\right|. \label{eq:afp}
\end{align}
{ It is still difficult to directly obtain the distance and angle sampling methods from (\ref{eq:afp}). 
However, it can be observed that the phase of each item in the summation of (\ref{eq:afp}) can be decoupled into two parts.} The first part is the linear phase $n\pi (\theta_q - \theta_p)$ related to the angle, while the second part is the 
 quadratic phase $k_cn^2d^2\left( \frac{1-\theta_p^2}{2r_p} - \frac{1-\theta_q^2}{2r_q}\right)$ related to the angle and distance.
Based on this observation, in the following two subsections, we will first derive the angular sampling method
from the first linear phase part, and then derive the distance sampling method
from the second quadratic phase part.

\subsection{Angular sampling method}
Firstly, to design the angular sampling method, we should focus on the first linear phase part  $n\pi (\theta_q - \theta_p)$ only related to the angle. 
For arbitrary two locations $(\theta_p, r_p)$ and $(\theta_q, r_q)$ in the polar coordinates, if $\frac{1-\theta_p^2}{r_p}$ and $\frac{1-\theta_q^2}{r_q}$ are equal to a constant $\frac{1}{\phi}$, i.e., $\frac{1-\theta_p^2}{r_p} = \frac{1-\theta_q^2}{r_q} = \frac{1}{\phi}$, then it is obvious that the quadratic phase part $k_cn^2d^2\left( \frac{1-\theta_p^2}{2r_p} - \frac{1-\theta_q^2}{2r_q}\right)$ becomes 0, and thus it can be removed. 
In this case, the column coherence $f(\theta_p, \theta_q,r_p,  r_q )$ is dependent on the first linear phase part  $n\pi (\theta_q - \theta_p)$, which only relies on the angles $\theta_p$ and $\theta_q$, while the effect of distance $r_p$ and $r_q$ is removed. Moreover, note that $(\theta_p, r_p)$, $(\theta_q, r_q)$ are two arbitrary locations satisfying $\frac{1-\theta_p^2}{r_p} = \frac{1-\theta_q^2}{r_q} = \frac{1}{\phi}$,  which means they are sampled from the curve $\frac{1-\theta^2}{r} = \frac{1}{\phi}$. For expression simplicity, we name the curve $\frac{1-\theta^2}{r} = \frac{1}{\phi}$ as distance ring ${\phi}$. As shown by the red curves in Fig. \ref{img:sampling} (b), different constants $\phi$ correspond to different distance rings ${\phi}$.
Thus, if the locations are sampled on the distance ring ${\phi}$, then the coherence $f(\theta_p, \theta_q, r_p, r_q)$ is only dependent on the angles $\theta_p$ and $\theta_q$. Then, we can derive the following angular sampling method.

Specifically, since $\frac{1-\theta_p^2}{r_p} = \frac{1-\theta_q^2}{r_q}$, from (\ref{eq:afp}), the coherence $f(\theta_p, \theta_q, r_p, r_q)$ can be expressed as 
\begin{align}
f(\theta_p, \theta_q, r_p, r_q) &= \left|\frac{1}{N}\sum_{n=-(N-1)/2}^{(N-1)/2}e^{jn\pi (\theta_q - \theta_p)  }\right| \notag\\&= \left|\frac{\sin(\frac{1}{2}N\pi(\theta_q - \theta_p))}{N\sin(\frac{1}{2}\pi(\theta_q - \theta_p))}\right|,\label{eq:theta}
\end{align}
which is only related to the angles $\theta_p$ and $\theta_q$. It can be found that (\ref{eq:theta}) is exactly equivalent to the coherence between two far-field steering vectors \cite{FundWC_Tse2015}. The zero points of the function (\ref{eq:theta}) satisfy $\theta_q - \theta_p = \frac{2m}{N}$, $m = 1,2,\cdots,N-1$.
Therefore, the angular sampling method on distance ring $\phi$ is the same as the existing angular sampling method of the angular-domain transform matrix $\mb{F}$.
In other words, angles should be uniformly sampled on distance ring $\phi$ as 
\begin{align}\label{eq:angle}
\theta_n = \frac{2n - N + 1}{N},\quad n = 0,1,\cdots, N-1.
\end{align}
\subsection{Distance sampling method}
Similar to the derivation of the angular sampling method, since the distance information is contained in the second quadratic phase part $k_cn^2d^2\left( \frac{1-\theta_p^2}{2r_p} - \frac{1-\theta_q^2}{2r_q}\right)$, we now focus on this part to derive the distance sampling method. 
For arbitrary two  vectors sampled on the same angle $\theta$, i.e., $\theta_p = \theta_q = \theta$, the first linear phase part $n\pi (\theta_q - \theta_p)$ becomes 0, thus being removed. 
In this case, the column coherence $f(\theta_p, \theta_q,r_p,  r_q ) = f(\theta, \theta,r_p,  r_q )$ is dependent on the second quadratic phase part  $k_cn^2d^2\left( \frac{1-\theta_p^2}{2r_p} - \frac{1-\theta_q^2}{2r_q}\right) =k_cn^2d^2\left( \frac{1-\theta^2}{2r_p} - \frac{1-\theta^2}{2r_q}\right)$, which relies on the distance-related items $\frac{1-\theta^2}{r_p}$ and $\frac{1-\theta^2}{r_q}$. 
The physical significance behind the discussion above is that, if two locations are sampled on the same angle $\theta$, the column coherence $f(\theta_p, \theta_q,r_p,  r_q )$ is only decided by the distance-related items $\frac{1-\theta^2}{r_p}$ and $\frac{1-\theta^2}{r_q}$. Based on this observation, we can derive the distance sampling method on the angle $\theta$.

Unfortunately, unlike the derivation of the angular sampling method, due to the quadratic phase property, it is difficult to get a close form like (\ref{eq:theta}). To cope with problem, in the following \textbf{Lemma 1}, the Fresnel functions are introduced to approximate the column coherence $f(\theta, \theta, r_p, r_q)$.
\begin{thm} \label{lemma1}
If two near-field steering vectors are sampled from the same angle $\theta$ but different distances $r_p$ and $r_q$, then the column coherence $f(\theta, \theta, r_p, r_q)$ can be approximated as 
\begin{align}
f(\theta, \theta, r_p, r_q)   \approx \left|G(\beta)\right| = \left|\frac{C(\beta) + jS(\beta)}{\beta}\right| 
,
\end{align}
where $\beta = \sqrt{\frac{N^2d^2(1 - \theta^2)}{2\lambda_c} \left| \frac{1}{r_p} - \frac{1}{r_q}\right|}$. $C(\beta) = \int_{0}^{\beta}\cos(\frac{\pi}{2}t^2)\text{d} t$ and $S(\beta) = \int_{0}^{\beta}\sin(\frac{\pi}{2}t^2)\text{d} t$ are Fresnel functions \cite{fresnel_Sherman1962}.
\end{thm}
\emph{Proof}: See Appendix A. $\hfill\blacksquare$
\begin{figure}	
	\centering
	\includegraphics[width=3.5in]{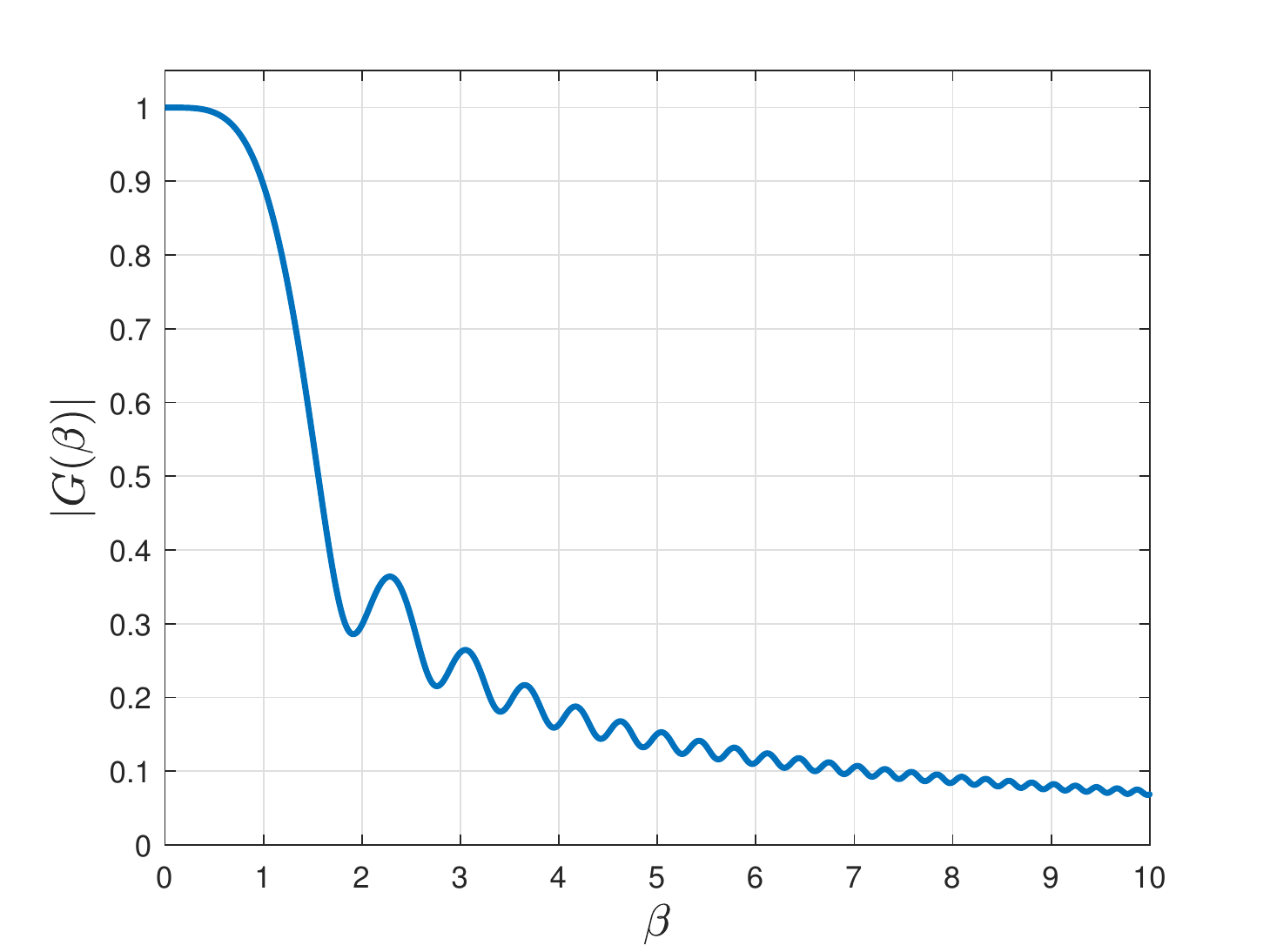}
	\vspace*{-1em}
	\caption{ The numerical results of $|G(\beta)|$ against $\beta$.
	}\label{img:gx}
	\vspace*{-1em}
\end{figure}

\textbf{Lemma 1} indicates that the column coherence heavily relies on the function $|G(\cdot)|$ and the parameter $\beta$. The function $|G(\cdot)|$ is composed of two Fresnel functions. Since the function $|G(\cdot)|$ does not contain any parameters, it is sufficient to obtain its numerical results through one numerical integration, and the result is shown in Fig. \ref{img:gx}. With the increase of $\beta$, $|G(\beta)|$ shows a significant downward trend with slight fluctuation. 
Therefore, in order to make the column coherence as small as possible, i.e., 
to let $f(\theta, \theta, r_p, r_q) \approx |G(\beta)|$  lower than a desired threshold $\Delta$, 
we should first calculate $\beta_{\Delta}$ satisfying $|G(\beta_\Delta)| = \Delta$. Then, due to the downward trend of the function $|G(\beta)|$, it can be approximately derived that $\beta \ge \beta_\Delta$. For example, if we desire that the column coherence is lower than $\Delta = 0.5$, then it can be solved from $|G(\beta_{0.5})| = 0.5$ that $\beta_{0.5} \approx 1.6$. Thus, we approximately have $\beta \ge 1.6$.

Based on the conditions $\beta \ge \beta_\Delta$ and $\beta = \sqrt{\frac{N^2d^2(1 - \theta^2)}{2\lambda_c} \left| \frac{1}{r_p} - \frac{1}{r_q}\right|}$, the sampled distances $r_p$ and $r_q$ should satisfy 
\begin{align}\label{eq:contraint1}
\left|\frac{1}{r_p} - \frac{1}{r_q}\right| \ge \frac{2\lambda_c\beta_\Delta^2}{N^2d^2(1-\theta^2)} = \frac{1}{Z_\Delta (1- \theta^2)},
\end{align}
where $Z_\Delta = \frac{N^2d^2}{2\lambda_c \beta_{\Delta}^2}$.  By considering the array aperture $D =  Nd$, we can rewrite $Z_{\Delta}$ as $Z_{\Delta} = \frac{D^2}{2\beta_{\Delta}^2\lambda_c}$.  We define $Z_{\Delta}$ as the threshold distance in this paper.

To make the column coherence lower than a given threshold, it is clear from (\ref{eq:contraint1}) that the difference between the inverses of two distances should be larger than a constant. 
For example, if we set $r_q = \frac{1}{s}Z_\Delta(1-\theta^2)$, $s = 1, 2, 3, \cdots$, then from (\ref{eq:contraint1}) we have 
\begin{align}\label{eq:contraint2}
r_p \ge \frac{1}{s-1}Z_\Delta(1-\theta^2)\quad \text{or}\quad r_p \le \frac{1}{s+1}Z_\Delta(1-\theta^2).
\end{align}
It can be inferred from (\ref{eq:contraint2}) that, on a arbitrary angle $\theta$, if one steering vector is sampled from the distance $\frac{1}{s}Z_\Delta(1-\theta^2)$, then to keep the coherence lower than $\Delta$, no other distances can be sampled in the range $\left[\frac{1}{s+1}Z_\Delta(1-\theta^2), \frac{1}{s-1}Z_\Delta(1-\theta^2)\right]$. In other words, $\frac{1}{s+1}Z_\Delta(1-\theta^2)$ is exactly the maximum feasible distance that is lower than $\frac{1}{s}Z_\Delta(1-\theta^2)$, while $\frac{1}{s-1}Z_\Delta(1-\theta^2)$ is exactly the minimum feasible distance that is higher than $\frac{1}{s}Z_\Delta(1-\theta^2)$.

Based on this observation, on the angle $\theta$, if the sampled distances are 
\begin{align} \label{eq:distance}
r_s = \frac{1}{s}Z_\Delta(1-\theta^2), \quad s = 0, 1,2,3,\cdots,
\end{align}
then the column coherence of two near-field steering vectors sampled at
two adjacent distances are exactly $\Delta$. 
That is to say, the constraint that the column coherence  is lower than $\Delta$ on any angle can be guaranteed.
Comparing (\ref{eq:angle}) and (\ref{eq:distance}), it can be found that the angle should be sampled uniformly, while the distance should be sampled non-uniformly.

\subsection{Design the polar-domain transform matrix}

In this subsection, based on the angular and distance sampling methods derived in (\ref{eq:angle}) and (\ref{eq:distance}), we conclude how to generate the polar-domain transform matrix $\mb{W}$ in \textbf{Algorithm 1}.
As shown in Fig. \ref{img:sampling} (b), the entire polar domain is divided into multiple distance rings $r_s$, while each distance ring is segmented by multiple angles. Therefore, the transform matrix $\mb{W} $ can be composed $S$ sub-matrices, i.e.,  
\begin{align}\label{eq:W}
\mb{W} = \left[\mb{W}_0, \mb{W}_2, \cdots, \mb{W}_{S-1} \right] \in \mathbb{C}^{N \times NS},
\end{align}
where $S$ denotes the number of distance rings, and the sub-matrix $\mb{W}_s \in \mathbb{C}^{N \times N}$ consists of $N$ near-field steering vectors sampled from the same curve $\frac{1 - \theta^2}{r} = \frac{1}{r_s}$ (or distance ring $r_s$) but different angles. Therefore, the number of column vectors in $\mb{W}$ is $Q = NS$.


\begin{algorithm}[htb]
	\caption{$\!\!$: The generating procedure of the proposed polar-domain transform matrix $\mb{W}$.}
	\label{alg:W}
	\begin{algorithmic}[1] 
		\REQUIRE ~~\\
		The minimum allowable distance $\rho_{\min}$; threshold $\beta_{\Delta}$; antenna number $N$; antenna spacing $d$; wavelength $\lambda_c$ \\
		\ENSURE ~~\\
		polar-domain transform matrix $\mb{W}$\\
		\STATE
		$Z_\Delta = \frac{N^2d^2}{2\beta_{\Delta}^2\lambda_c}$\\
		\STATE
		$s = 0$ \\
		\REPEAT
		\FOR{$n\in\{0,1,\cdots,N-1\}$}		
		\STATE 
		$\theta_n = \frac{2n-N + 1}{N}$ according to (\ref{eq:angle})\\
		\STATE 
		$r_{s, n} = \frac{1}{s}Z_\Delta(1 - \theta_n^2)$ according to (\ref{eq:distance})\\
		\ENDFOR
		\STATE
		$\mb{W}_s = [\mb{b}(\theta_0, r_{s, 0 }),\mb{b}(\theta_1, r_{s,1}),\cdots,\mb{b}(\theta_{N - 1}, r_{s,N - 1}) ]$\\
		\STATE
		$S = s, \:\: s = s + 1 $
		\UNTIL $ \frac{1}{s}Z_\Delta < \rho_{\min} $	
		\STATE $\mb{W}  = [{\mb{W}}_1,{\mb{W}}_2,\cdots, {\mb{W}}_{S}]$
		\RETURN $\mb{W} $.
	\end{algorithmic}
\end{algorithm}
Specifically, in step 1 of \textbf{Algorithm 1}, the threshold distance $Z_\Delta$ is calculated. Next, in steps 3-10, different sub-matrices $\mb{W}_s$ are sequentially generated. 
 In steps 5-6, $N$ angles and distances are sampled according to (\ref{eq:angle}) and (\ref{eq:distance}).
 After that, in step 8, the sub-matrix $\mb{W}_s$ is generated based on the sampled angles and distances. 
 It can be derived from steps 5-6 that $\theta_n$ and $r_{s,n}$ are sampled on the curve $\frac{1-\theta_n^2}{r_{s, n}} = \frac{1-\theta_n^2}{\frac{1}{s}Z_\Delta(1 - \theta_n^2)} = \frac{s}{Z_\Delta}$, which is exactly equivalent to the distance ring $r_s = 1/\frac{s}{Z_\Delta} = \frac{1}{s}Z_\Delta$. 
 Thus the $s$-th sub-matrices $\mb{W}_s$ is sampled from the distance ring $r_s$. 
 Furthermore, in step 10, since the actual distances between the BS and the users or scatters cannot be zero, we define $\rho_{\text {min}}$ as the minimum allowable distance ring. Then, only when the distance ring $r_s = \frac{1}{s}Z_\Delta$ is larger than $\rho_{\text {min}}$ can it be sampled, just as shown in step 3. Thus, the number of distances rings $S$ is determined by $\rho_{\text{min}}$.
 Finally, in step 11, all sub-matrices are concatenated to construct the polar-domain transform matrix $\mb{W}$.

It is worth noting that since the polar-domain transform matrix $\mb{W}$ also samples distances in the far-field region, e.g., $s = 0$ (or $r_s = +\infty$)  in (\ref{eq:distance}), when the minimum allowable distance $\rho_{\text{min}}$ is large enough, only the distances in the far-field will be sampled. Then, the polar-domain representation becomes the angular-domain representation, and therefore the angular-domain representation is a special case of the proposed polar-domain representation.

Based on the proposed polar-domain representation of the near-field XL-MIMO channel, we will propose the on-grid and  off-grid near-field channel estimation schemes in the next section.

	\section{Proposed Near-Field Channel Estimation Schemes} \label{sec:4}
In this section, by exploiting the proposed polar-domain sparse representation of the near-field channel, we first propose an on-grid near-field channel estimation algorithm called polar-domain simultaneous orthogonal matching pursuit (P-SOMP) to efficiently estimate the near-field  XL-MIMO channel. Then, an off-grid near-field channel estimation algorithm called polar-domain simultaneous iterative gridless weighted (P-SIGW) is further proposed to improve the channel estimation accuracy.

\subsection{On-grid near-field channel estimation}
{As we discussed in section \ref{sec:2}, since the $K$ users transmit mutual orthogonal pilot sequences, uplink channel estimation for each user can be carried out independently.} For an arbitrary user, based on the polar-domain representation (\ref{eq:pt}), the received pilot $\mb{y}_m$ at frequency $f_m$ in (\ref{eq:HP}) can be represented as 
	\begin{align} \label{eq:PF}
	\mb{y}_m = \mb{A} \mb{W}\mb{h}_m^{\mathcal{P}} + \mb{n}_m = \mb{\Psi} \mb{h}_m^{\mathcal{P}} + \mb{n}_m,
	\end{align}
where $\mb{\Psi} = \mb{A}\mb{W}$. Under the CS framework \cite{CS_Ba10}, each element of $\mb{A}$ can be randomly selected from $\frac{1}{\sqrt{N}}\{-1, +1\}$ with equal probability. Since the polar-domain channel $\mb{h}_m^P$ is sparse as discussed in Section \ref{sec:3}, the polar-domain channel estimation can be formulated as a sparse signal recovery problem. 

However, since the received noise $\mb{n}_m = \left[ \mb{n}^T_{m,1} \mb{A}^T_{1}, \cdots, \mb{n}^T_{m,P}\mb{A}^T_{P} \right]^T$ is colored noise, a pre-whitening procedure should be carrier out at first \cite{SWOMP_Robert18}. To be specific, the covariance matrix of the noise is $\mb{C} = \mathbb{E}\left(\mb{n}_m \mb{n}_m^H\right) = \text{blkdiag}\left(\sigma^2\mb{A}_1\mb{A}_1^H, \sigma^2\mb{A}_2\mb{A}_2^H,\cdots, \sigma^2\mb{A}_P\mb{A}_P^H\right)$. Then, this covariance matrix can be decomposed by Cholesky factorization as $\mb{C} = \sigma^2\mb{D}\mb{D}^H$, where $\mb{D} \in \mathbb{C}^{PN_{\text{RF}} \times PN_{\text{RF}}}$ is a lower triangular matrix. Thus the pre-whitening matrix is $\mb{D}^{-1}$, and then the whitened received signal $\bar{\mb{y}}_m$ at frequency $f_m$ is
\begin{align}\label{eq:PF2}
\bar{\mb{y}}_m=\mb{D}^{-1}\mb{y}_m=\bar{\mb{\Psi}}\mb{h}_m^{\mathcal{P}} + \bar{\mb{n}}_m,
\end{align}
where $\bar{\mb{\Psi}} =\mb{D}^{-1}\mb{A}\mb{W}$ and $\bar{\mb{n}}_m = \mb{D}^{-1}\mb{n}_m$. In this case, the covariance matrix of $\bar{\mb{n}}_m$ is $\bar{\mb{C}} = \mb{D}^{-1}\mb{C}\mb{D}^{-H} = \sigma^2 \mb{I}_{PN_{\text{RF}}}$, thus the noise $\bar{\mb{n}}_m$ becomes white.

 Generally, the steering vectors at different sub-carriers are the same \cite{CE_SOMP_Gao16}, just as the frequency-independent steering vector $\mb{b}(\cdot)$ and $\mb{a}(\cdot)$. Therefore, the sparsity support of the polar-domain channels $\mb{h}_m^P$ at different subcarriers $f_m$ are also the same, and they can be simultaneously  estimated to increase the estimation accuracy. 
 Therefore, we rearrange (\ref{eq:PF2}) as 
\begin{align} \label{eq:PF3}
	\bar{\mb{Y}} =  \mb{D}^{-1}{\mb{Y}} = \bar{\mb{\Psi}}\mb{H}^{\mathcal{P}} + \bar{\mb{N}},
\end{align}
where $\bar{\mb{Y}} = [\bar{\mb{y}}_1, \bar{\mb{y}}_2, \cdots, \bar{\mb{y}}_M]$, $\mb{Y} = [\mb{y}_1, \cdots, \mb{y}_M]$, 
 $\mb{H}^{\mathcal{P}} = [\mb{h}_1^{\mathcal{P}}, \cdots, \mb{h}_M^{\mathcal{P}}]$, and $\bar{\mb{N}} = [\bar{\mb{n}}_1, \cdots, \bar{\mb{n}}_M]$. The target is to estimate the channel $\mb{H} = \mb{W}\mb{H}^{\mathcal{P}}$ from $\bar{\mb{\Psi}}$ and $\bar{\mb{Y}}$.  Leveraging the channel sparsity in the polar domain, the row of $\mb{H}^{\mathcal{P}}$ is sparse. Therefore, the channel estimation problem can be solved by the existing simultaneous orthogonal matching pursuit (SOMP) algorithm \cite{SWOMP_Robert18}.
 
 In this paper, we extend the classical angular-domain SOMP algorithm to a polar-domain SOMP (P-SOMP) algorithm to recover the near-field XL-MIMO channel.  
 The proposed P-SOMP algorithm  is illustrated in \textbf{Algorithm 2}. 
\begin{algorithm}[htb]
	\caption{$\!\!$: The proposed polar-domain SOMP algorithm.}
	\label{alg:PSOMP}
	\begin{algorithmic}[1]
		\REQUIRE ~~\\
		Received pilot $\mb{Y}$; combining matrix $\mb{A}$; the minimum distance $\rho_{\min}$; number of paths $\hat{L}$\\
		\ENSURE ~~\\
		The estimated near-field channel $\hat{\mb{H}}$\\
		\STATE 
		Construct the polar-domain transform matrix $\mb{W}$ by \textbf{Algorithm 1}
		\STATE
		Covariance matrix $\mb{C} = \text{blkdiag}\left(\mb{A}_1\mb{A}_1^H, \cdots, \mb{A}_P\mb{A}_P^H\right)$
		\STATE 
		Calculate the pre-whitening matrix $\mb{D}$ by solving $\mb{C} = \mb{D}\mb{D}^H$
		\STATE 
		Pre-whitening: $\bar{\mb{Y}} = \mb{D}^{-1}\mb{Y}$, $\bar{\mb{\Psi}} =\mb{D}^{-1}\mb{A}\mb{W}$\\
		\STATE 
		Initialization: $\mb{R} = \bar{\mb{Y}}$, $\varUpsilon = \{ \emptyset \}$\\
		\FOR {$l \in \{1,2,\cdots,\hat{L}\}$}
		\STATE 
		Calculate the correlation matrix: $\mb{\Gamma} = \bar{\mb{\Psi}}^H \mb{R}$\\
		\STATE
		Detect new support: $p^{\star} = \arg \max_p \sum_{m = 1}^{M} |\Gamma(p, m)|^2$\\
		\STATE 
		Update support set: $\varUpsilon = \varUpsilon \cup p^{\star}$\\
		\STATE 
		Orthogonal projection: $\hat{\mb{H}}^{\mathcal{P}}_{\varUpsilon, :} = \bar{\mb{\Psi}}_{:,\varUpsilon}^{\dag}\bar{\mb{Y}}$\\	
		\STATE 
		Update residual: $\mb{R} = \mb{R} - \bar{\mb{\Psi}}_{:,\varUpsilon}\hat{\mb{H}}^{\mathcal{P}}_{\varUpsilon, :} $\\
		\ENDFOR
		\STATE 
		$\hat{\mb{H}} = \mb{W}_{:,\varUpsilon}\hat{\mb{H}}^{\mathcal{P}}_{\varUpsilon, :}$\\
		\RETURN $\hat{\mb{H}} $.
	\end{algorithmic}
\end{algorithm}

Specifically, in step 1, we first construct the polar-domain transform matrix $\mb{W}$ according to $\textbf{Algorithm 1}$. Next in steps 2-4, the pre-whitening procedure is carried out to whiten the received signal. 
Then, in steps 5-12, we utilize the SOMP algorithm to successively estimate the physical channel angle and distance in  the polar domain. 
In step 5, we initialize the residual matrix $\mb{R} = \bar{\mb{Y}}$ and the sparse support set $\varUpsilon = \{ \emptyset \}$. 
Then, for the $l$-th path component, we first calculate the correlation matrix $\mb{\Gamma} = \bar{\mb{\Psi}}^H\mb{R}$ in step 7. Next in steps 8, based on the assumption that the support sets at different subcarriers are the same, the power of the correlation matrix on the $p$-th row is $\sum_{m = 1}^{M} |\Gamma(p, m)|$. 
Thus, the index $p^{\star}$ of the physical location of the $l$-th path component can be determined as $p^{\star} = \arg \max_p \sum_{m = 1}^{M} |\Gamma(p, m)|$.
Then, we add $p^{\star}$ to the sparse support set $\varUpsilon$. After that, in step 10, the path gain $\hat{\mb{H}}^{\mathcal{P}}_{\varUpsilon, :}$ on the support set $\varUpsilon$ is calculated through orthogonal least square, and in step 11, we update the residual matrix $\mb{R}$. The steps above are carried out for $\hat{L}$ times until all path components are detected. Finally, the near-field channel $\hat{\mb{H}}$ is  recovered as 	$\hat{\mb{H}} = \mb{W}_{:,\varUpsilon}\hat{\mb{H}}^{\mathcal{P}}_{\varUpsilon, :}$.

The main difference between the proposed P-SOMP algorithm and the existing SOMP algorithm is that, the proposed P-SOMP algorithm is carried out in the polar domain, so it can efficiently estimate the near-field XL-MIMO channel. Moreover, since the polar-domain transform matrix $\mb{W}$ also samples distances in the far-field region, e.g., $s = 0$ in (\ref{eq:distance}), the P-SOMP algorithm also works well in the far-field, which will be verified by simulations in Section \ref{sec:5}.

However, the proposed P-SOMP algorithm assumes that the angles and distances exactly lie in the sampled points in the polar domain, i.e., on-grid angles and distances. In contrast, the actual angles and distances are continuously distributed, i.e., off-grid angles and distances. Then, the estimation accuracy of the proposed on-grid P-SOMP algorithm is limited, which will be improved in the next Subsection \ref{sec:4-2}.

\subsection{Off-grid near-field channel estimation}\label{sec:4-2}
 To cope with the estimation error introduced by the on-grid sample points, inspired by the classical off-grid simultaneous gridless weighted (SIGW) algorithm in the angular-domain, we propose an off-grid near-field channel estimation algorithm called polar-domain simultaneous gridless weighted (P-SIGW) to improve the channel estimation performance.
 Unlike the existing SIGW algorithm that only refines the estimated path gains and angles, the proposed P-SIGW algorithm simultaneously refines the path gains, angles, and distances by following the maximum likelihood principle.
Specifically, the proposed P-SIGW algorithm is provided in \textbf{Algorithm 3}.

	\begin{algorithm}[htb]
	\caption{$\!\!$: The proposed polar-domain SIGW algorithm.}
	\label{alg:PSIGW}
	\begin{algorithmic}[1]
		\REQUIRE ~~\\
		Received pilot sequences $\mb{Y}$; combining matrix $\mb{A}$; the minimum distance $\rho_{\min}$; number of detected paths $\hat{L}$, number of iterations $N_\text{iter}$\\
		\ENSURE ~~\\
		The estimated near-field channel $\hat{\mb{H}}$\\
		\textbf{Initialization stage}
		\STATE 
		Obtain the initial value of the distances $\hat{\mb{r}}^0 = [\hat{r}_1^0, \hat{r}_2^0, \cdots, \hat{r}_{\hat{L}}^0]$ and the angles $\hat{\bm{\theta}}^0 = [\hat{\theta}_1^0, \hat{\theta}_2^0, \cdots, \hat{\theta}_{\hat{L}}^0]$ by \textbf{Algorithm 2}.\\
		\textbf{Refinement stage}
		\FOR {$n \in \{1,2,\cdots,N_\text{iter}\}$}
		\STATE
		Choose Armijo backtracking line search step length $l_1$\\
		\STATE 
		Update the angles by $\hat{\bm{\theta}}^{n} = \hat{\bm{\theta}}^{n-1} - l_1\nabla_{\hat{\bm{\theta}}}\mathcal{L}(\hat{\bm{\theta}}, \hat{\mb{r}}^{n-1})|_{\hat{\bm\theta} = \hat{\bm \theta}^{n-1} }$ by (\ref{eq:thetan}) \\
				\STATE
		Choose Armijo backtracking line search step length $l_2$\\
		\STATE 
		Update the distances by $\frac{1}{\hat{\mb{r}}^{n}} =  \frac{1}{\hat{\mb{r}}^{n-1}} - l_2\nabla_{\frac{1}{\hat{\mb{r}}}}\mathcal{L}(\hat{\bm\theta}^{n}, \hat{\mb{r}})|_{\hat{\mb{r}} = \hat{\mb{r}}^{n-1}}$ by (\ref{eq:rn})\\
		\STATE
		Update the path gains $\hat{\mb{G}}^{n}$ by (\ref{eq:G})
		\ENDFOR
		\STATE 
		$\hat{\mb{H}} = [\mb{b}(\hat{\theta}_1^n, \hat{r}_1^n), \mb{b}(\hat{\theta}_2^n, \hat{r}_2^n),\cdots, \mb{b}(\hat{\theta}_{\hat L}^n, \hat{r}_{\hat L}^n)]\hat{\mb{G}}^n$\\
		\RETURN $\hat{\mb{H}} $.
	\end{algorithmic}
\end{algorithm}

The proposed P-SIGW algorithm is composed of an initialization stage and a refinement stage.
Firstly, in step 1, we regard the P-SOMP algorithm as an initialization stage of the P-SIGW algorithm. After carrying out \textbf{Algorithm 2}, we obtain the initial value of the estimated distances $\hat{\mb{r}} = [\hat{r}_1, \hat{r}_2,\cdots, \hat{r}_{\hat{L}}]$, angles $\hat{\bm{\theta}} = [\hat{\theta}_1, \hat{\theta}_2,\cdots, \hat{\theta}_{\hat{L}}]$, and complex path gains $\hat{\mb{G}} = \hat{\mb{H}}^P_{\varUpsilon,:} \in \mathbb{C}^{\hat{L} \times M}$. 

Then, in the refinement stage, we concatenate the detected near-field paths to construct the matrix $\tilde{\mb{W}}(\hat{\bm{\theta}}, \hat{\mb{r}}) = [\mb{b}(\hat{\theta}_1, \hat{r}_1), \mb{b}(\hat{\theta}_2, \hat{r}_2),\cdots, \mb{b}(\hat{\theta}_{\hat L}, \hat{r}_{\hat L})]$, thus the recovered channel can be written as $\hat{\mb{H}} = \tilde{\mb{W}}(\hat{\bm{\theta}}, \hat{\mb{r}})\hat{\mb{G}}$. In steps 2-7, the distances $\hat{\mb{r}}$, angles $\hat{\bm{\theta}}$, and path gains $\hat{\mb{G}}$ are alternatively optimized to maximize the likelihood, which is formulated as  
\begin{align} \label{eq:PF4}
	\min_{\hat{\mb{G}}, \hat{\bm{\theta}}, \hat{\mb{r}}} \|\bar{\mb{Y}} - \tilde{\mb{\Psi}}(\hat{\bm{\theta}}, \hat{\mb{r}})\hat{\mb{G}} \|_F^2,
\end{align}
where $\tilde{\mb{\Psi}}(\hat{\bm{\theta}}, \hat{\mb{r}}) = \mb{D}^{-1}\mb{A}\tilde{\mb{W}}(\hat{\bm{\theta}}, \hat{\mb{r}})$.
Since the optimization problem (\ref{eq:PF4}) is non-convex, we utilize the alternating minimization method to solve this problem. 
For fixed $\hat{\mb{r}}$ and $\hat{\bm{\theta}}$, the optimal solution for $\hat{\mb{G}}$ is given by
\begin{align} \label{eq:G}
	\hat{\mb{G}}^{\text{opt}} = \tilde{\mb{\Psi}}^{\dagger}(\hat{\bm{\theta}}, \hat{\mb{r}})\bar{\mb{Y}}.
\end{align}
Substitute (\ref{eq:G}) into (\ref{eq:PF4}), the maximum-likelihood problem is reformulated as 
\begin{align} \label{eq:PF5}
&\min_{\hat{\bm{\theta}}, \hat{\mb{r}}}  \|\bar{\mb{Y}} - \tilde{\mb{\Psi}}(\hat{\bm{\theta}}, \hat{\mb{r}})\tilde{\mb{\Psi}}^{\dagger}(\hat{\bm{\theta}}, \hat{\mb{r}})\bar{\mb{Y}} \|_F^2 
\notag\\\Leftrightarrow \: &\min_{\hat{\bm{\theta}}, \hat{\mb{r}}}  \text{Tr}\left\{\bar{\mb{Y}}^H\left(\mb{I} - \mb{P}(\hat{\bm{\theta}}, \hat{\mb{r}})\right)^H\left(\mb{I} - \mb{P}(\hat{\bm{\theta}}, \hat{\mb{r}})\right)\bar{\mb{Y}}\right\} \notag \\
\overset{ (a) }{\Leftrightarrow} \: &\min_{\hat{\bm{\theta}}, \hat{\mb{r}}}\mathcal{L}(\hat{\bm{\theta}}, \hat{\mb{r}}) = -\text{Tr}\left\{\bar{\mb{Y}}^H\mb{P}(\hat{\bm{\theta}}, \hat{\mb{r}})\bar{\mb{Y}}\right\},
\end{align}
where $\mb{P}(\hat{\bm{\theta}}, \hat{\mb{r}}) = \tilde{\mb{\Psi}}(\hat{\bm{\theta}}, \hat{\mb{r}})\tilde{\mb{\Psi}}^{\dagger}(\hat{\bm{\theta}}, \hat{\mb{r}})$, and (a) is derived by $\mb{P}^H(\hat{\bm{\theta}}, \hat{\mb{r}})\mb{P}(\hat{\bm{\theta}}, \hat{\mb{r}}) = \mb{P}(\hat{\bm{\theta}}, \hat{\mb{r}})$.  The new objective function $\mathcal{L}(\hat{\bm{\theta}}, \hat{\mb{r}})$ can be optimized using an iterative gradient descent approach. In the $n$-th iteration, the angles are updated as 
	\begin{align} 
	 \hat{\bm{\theta}}^{n} &= \hat{\bm{\theta}}^{n-1} - l_1\nabla_{\hat{\bm{\theta}}}\mathcal{L}(\hat{\bm{\theta}}, \hat{\mb{r}}^{n-1})|_{\hat{\bm\theta} = \hat{\bm \theta}^{n-1} }, \label{eq:thetan}
	\end{align}
where $l_1$ denotes the step length for the angles. Moreover, as for the distance, we found that if the gradient with respect to $\hat{\mb{r}}$ is directly utilized to update $\mb{r}$, the estimation performance fluctuates with distance. This fact can be explained by the distance-sampling method (\ref{eq:distance}) that $r$ is non-uniformly sampled, or in other words, $\frac{1}{r}$ is uniformly sampled. Motivated by this observation, we define $\frac{1}{\hat{\mb{r}}} = [\frac{1}{\hat{r}_1}, \frac{1}{\hat{r}_2},\cdots, \frac{1}{\hat{r}_{\hat{L}}}]$, and utilize the gradient with respect to $\frac{1}{\hat{\mb{r}}}$ to indirectly update $\hat{\mb{r}}$, i.e., 
\begin{align} 
\frac{1}{\hat{\mb{r}}^{n}} &=  \frac{1}{\hat{\mb{r}}^{n-1}} - l_2\nabla_{\frac{1}{\hat{\mb{r}}}}\mathcal{L}(\hat{\bm\theta}^{n}, \hat{\mb{r}})|_{\hat{\mb{r}} = \hat{\mb{r}}^{n-1}}\label{eq:rn},
\end{align}
where $l_2$ denotes the step length for the inverse of distances. {To guarantee the objective function is non-increasing, 
the step lengths $l_1$ and $l_2$ are chosen by Armijo backtracking line search.} 
The gradient $\nabla\mathcal{L}(\hat{\bm{\theta}}, \hat{\mb{r}})$ is given in Appendix B. 
Based on (\ref{eq:G}), (\ref{eq:thetan}), and (\ref{eq:rn}), the parameters are updated in steps 3-7 in \textbf{Algorithm 3}. 
Finally, all refined paths are concatenated to reconstruct the near-field channel $\hat{\mb{H}}$. 
In section \ref{sec:5}, simulation results will be provided to verify the effectiveness of the proposed on-grid P-SOMP algorithm and off-grid P-SIGW algorithm.

{
\subsection{Complexity and convergence analysis}\label{sec:4-3}	
\emph{Convergence}: The convergence of the proposed optimization algorithm P-SIGM is discussed here. Since the objective function $ \|\bar{\mb{Y}} - \tilde{\mb{\Psi}}(\hat{\bm{\theta}}, \hat{\mb{r}})\hat{\mb{G}} \|_F^2$ is larger than 0, it has a lower-bound. Then in each iteration, $\hat{\mb{G}}^{\text{opt}}$ is the optimal solution to update $\hat{\mb{G}}$. Moreover, for the steps of updating $\hat{\bm{\theta}}$ and $\hat{\mb{r}}$, their step lengths $l_1$ and $l_2$ are chosen by Armijo backtracking line search. 
		Therefore, the steps for updating $\hat{\mb{G}}$, $\hat{\bm{\theta}}$, and $\hat{\mb{r}}$   are monotonically non-increasing, and the alternating procedure will converge.
\footnote{Notice that the proposed optimization algorithm converges to a feasible
	solution, while only the convergence rate is fully analyzed can its local optimality
	be strictly proved \cite{AO_Bezdek2002, AO_Bezdek2003}.}

		\begin{table}[!t]
	\caption{Computational complexity comparison}
	\label{table1}
	\centering
	\begin{tabular}{|c|c|}
		\hline
				Algorithm            &  Computational complexity $\mathcal{O}(\cdot)$     \\ \hline 
		P-SOMP            &  $\mathcal{O}(\hat{L}PN_{\text{RF}}NSM)$     \\ \hline                                                                                         
		SWOMP              &  $\mathcal{O}(\hat{L}PN_{\text{RF}}NM)$ \\ \hline                 
		P-SIGW   &  $\mathcal{O}(\hat{L}PN_{\text{RF}}NSM) +  \mathcal{O}( N_{\text{iter}}(P^2N_{\text{RF}}^2M + PN_{\text{RF}}M^2 )) $ \\ \hline
		SS-SIGW-OLS     & $\mathcal{O}(\hat{L}PN_{\text{RF}}NM) + \mathcal{O}( N_{\text{iter}}(P^2N_{\text{RF}}^2M + PN_{\text{RF}}M^2 )) $\\ \hline
	\end{tabular}
\vspace*{-2em}
\end{table}

\emph{Complexity}: The computational complexities of the proposed P-SOMP and P-SIGW schemes are analyzed, which are summarized in Table \ref{table1}.
For \textbf{Algorithm 2}, the overall complexity is mainly determined by operations of the SOMP procedure, i.e., the steps 6-12 in \textbf{Algorithm 2}. Thus we only consider these steps. Since $\bar{\mb{\Psi}}\in \mathbb{C}^{PN_{RF} \times NS}$, $\mb{R} \in \mathbb{C}^{PN_{RF} \times M}$, and $\mb{Y} \in \mathbb{C}^{PN_{RF} \times M}$, the computation complexities from step 7-11 are $\mathcal{O}(PN_{\text{RF}}NSM)$,  	 $\mathcal{O}(NSM)$,  	$\mathcal{O}(1)$,  	$\mathcal{O}( \hat{L}^2PN_{\text{RF}} +  \hat{L}PN_{\text{RF}}M)$,  and	$\mathcal{O}( \hat{L}PN_{\text{RF}}M)$,  respectively. Generally, the number of paths $\hat{L}$ is small, thus the computation complexity from steps 7-11 is decided by $\mathcal{O}(PN_{\text{RF}}NSM)$. After $\hat{L}$ times iterations, the overall complexity is $\mathcal{O}(\hat{L}PN_{\text{RF}}NSM)$. As shown in Table \ref{table1}, the computation complexity of the proposed P-SOMP is $S$ times that of the far-field on-grid scheme SWOMP \cite{SWOMP_Robert18}. However, the number of sampled distance rings $S$ is generally very small. For instance, in our simulations, $S$ is chosen from 4 to 6. Thus, this computation complexity is acceptable.

For \textbf{Algorithm 3}, its complexity is determined by the initialization stage and the refinement stage. The complexity in the initialization stage is the same as that of \textbf{Algorithm 2}, i.e., $\mathcal{O}(\hat{L}PN_{\text{RF}}NSM)$. Moreover, the complexity in the refinement stage is mainly introduced by the updates of the variables $\hat{\bm{\theta}}$, $\hat{\mb{r}}$, and $\hat{\mb{G}}$. As shown in (\ref{eq:G}), the complexity to update $\hat{\mb{G}}$ is $\mathcal{O}( \hat{L}^2PN_{\text{RF}} +  \hat{L}PN_{\text{RF}}M)$. As shown in Appendix B, the complexities of (\ref{eq:ap2_1}),  (\ref{eq:ap2_4}), (\ref{eq:ap2_5}), and (\ref{eq:ap2_6}) to calculate the gradients with respect to $\hat{\bm{\theta}}$ or $\hat{\mb{r}}$ are  $\mathcal{O}( P^2N_{\text{RF}}^2M + PN_{\text{RF}}M^2) $,  $\mathcal{O}( \hat{L}^2PN_{\text{RF}} + \hat{L}(PN_{\text{RF}})^2) $, $\mathcal{O}( \hat{L}^3 + \hat{L}^2PN_{\text{RF}}) $, and $\mathcal{O}( PN_{\text{RF}}N) $. Since $\hat{L}$ is very small, the complexity for each iteration in the refinement stage is determined by $\mathcal{O}( P^2N_{\text{RF}}^2M + PN_{\text{RF}}M^2) $.  After $N_{\text{iter}}$ times iterations, the complexity for refinement becomes $\mathcal{O}( N_{\text{iter}}(P^2N_{\text{RF}}^2M + PN_{\text{RF}}M^2 )) $. In conclusion, the overall complexity is  $\mathcal{O}(\hat{L}PN_{\text{RF}}NSM + N_{\text{iter}}(P^2N_{\text{RF}}^2M + PN_{\text{RF}}M^2)) $. It can be observed from Table \ref{table1} that, the complexity for refinement of the proposed P-SIGW scheme is the same as that of the far-field off-grid SS-SIGW-OLS \cite{WideCT_Go21}. The complexity difference between these two algorithms is introduced by their initialization stages, which has been discussed before.
}	
	\section{Simulation Results} \label{sec:5}

		In this section, simulation results are provided to verify the performance of the proposed near-field channel estimation schemes. The performance is evaluated by the normalized mean square error (NMSE), which is  defined as 		$
		\text{NMSE} = \mathbb{E}\left( \frac{\|\mb{H} - \hat{\mb{H}}\|_2^2 }{\|\mb{H} \|_2^2} \right)
		$.
		{ A multi-user XL-MIMO OFDM system is considered, and the simulation configurations are shown in Table \ref{table2}.}

		\begin{table}[!t]
			\caption{Simulation Configurations}
			\label{table2}
			\centering
			\begin{tabular}{|c|c|}
				\hline
				The number of BS antennas $N$            &  256     \\ \hline                                                                                         
				The number of Users    $K$               &  4 \\ \hline                 
				The number of RF chains  $N_{\text{RF}}$ &  4 \\ \hline
				Carrier frequency    $f_c$ & 100 GHz \\ \hline
				Bandwidth $B$ & 100 MHz \\ \hline
				The number of subcarriers $M$ & 256 \\ \hline
				The minimum allowable distance $\rho_{\text{min}}$ & 3 meters \\ \hline
				The number of channel paths $L$ per user & 6 \\ \hline 
				The distribution of $\theta$ &  $\mathcal{U}\left(- \frac{\sqrt{3}}{2} , \frac{\sqrt{3}}{2}\right)$ \\ \hline 
				Signal-to-noise ratio SNR & $1 / \sigma^2$ \\ \hline
				Parameter $\beta_\Delta$ & 1.2 \\ \hline 
				The number of iterations $N_{\text{iter}}$ & 10 \\ \hline
				The number of paths to be detected $\hat{L}$ & 12 \\ \hline
			\end{tabular}
		\vspace*{-2em}
		\end{table}
		
		\begin{figure}
			\centering
			\includegraphics[width=3.5in]{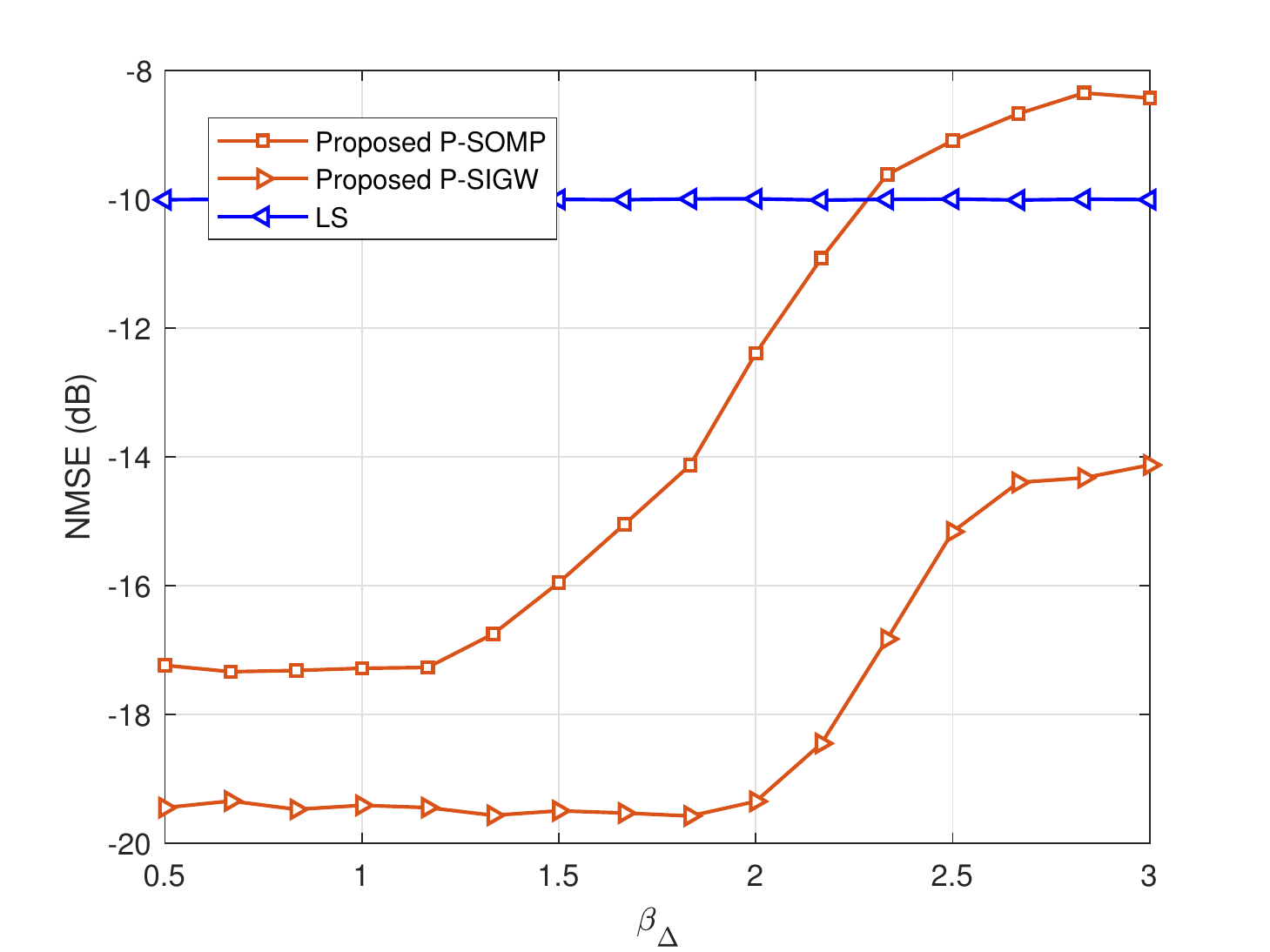}
			\vspace*{-1em}
			\caption{ NMSE performance against the parameter $\beta_\Delta$.
			}
			\label{img:nmse_beta}
			\vspace*{-1em}
		\end{figure}
		Firstly, to determine the value of parameter $\beta_\Delta$, the algorithm performance with respect to $\beta_\Delta$ is evaluated in Fig. \ref{img:nmse_beta}. The distances between the BS and the users or scatters are chosen from  $\mathcal{U}(5$ m$, 10$ m$)$, the pilot length is 32, and the SNR is 10 dB. The proposed on-grid scheme will not suffer from a severe performance loss if $\beta_\Delta < 1.2$, while the proposed off-grid scheme will not suffer from a severe performance loss if $\beta_\Delta < 2$. Therefore, to guarantee the performance of the proposed two algorithms, we set $\beta_\Delta = 1.2$ in the following simulations. Under such settings, the number of sampled distances on each angle is $S = 6$.

		\begin{figure}
			\centering
			{\includegraphics[width=3.5in]{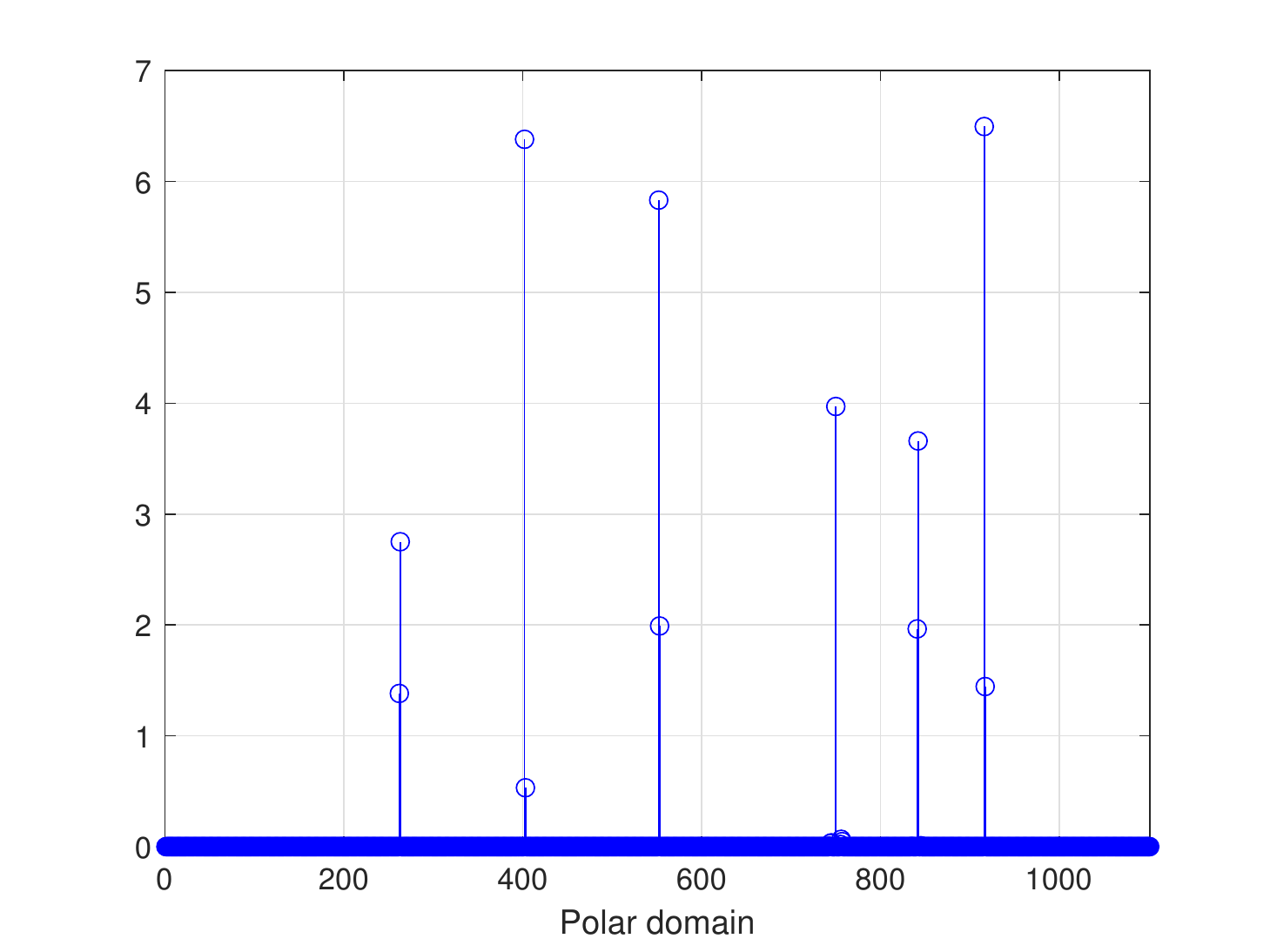}} 
			\\ 
			\centering
			\vspace*{-1em}
			\caption{The proposed polar domain channel representation. 
			} 
			\label{img:sparse_representation}
		\end{figure}
		
		In Fig. \ref{img:sparse_representation}, the polar-domain channel in near-field is plotted to show the polar-domain sparsity. The distances between the BS and the users or scatters are randomly chosen from $\mathcal{U}(5$ m$, 10$ m$)$. Since the angular-domain transform matrix $\mb{F} \in \mathbb{C}^{N \times N}$ is an orthogonal matrix, the angular-domain channel $\mb{h}_m^{\mathcal{A}}$ can be obtained as $\mb{h}_m^{\mathcal{A}} = \mb{F}^H \mb{h}_m$. 
		However, as for the polar-domain transform matrix $\mb{W} \in \mathbb{C}^{N \times NS} $, since the number of columns $NS$ is larger than $N$, the polar-domain channel $\mb{h}_m^{\mathcal{P}}$ can not be directly obtained by $\mb{h}_m^{\mathcal{P}} = \mb{W}^H \mb{h}_m$. To explicitly illustrate $\mb{h}_m^{\mathcal{P}}$, we utilize compressed sensing methods to obtain $\hat{\mb{h}}_m^{\mathcal{P}}$ from ${\mb{h}}_m$ by solving $\mb{h}_m=\mb{W}\mb{h}_m^{\mathcal{P}}$, which is plotted in Fig. \ref{img:sparse_representation}. In our simulations, we found that the NMSE between  $\hat{\mb{h}}_m = \mb{W}\hat{\mb{h}}_m^{\mathcal{P}}$ and  $\mb{h}_m$ is lower than -20 dB, so we can assume 
		$\hat{\mb{h}}_m^{\mathcal{P}} \approx \mb{h}_m^{\mathcal{P}}$.
		The number of paths $L$ is 6, and there are exactly 6 peaks in the polar-domain channel, which shows obvious sparsity.

	\begin{figure}
	\centering
	\includegraphics[width=3.5in]{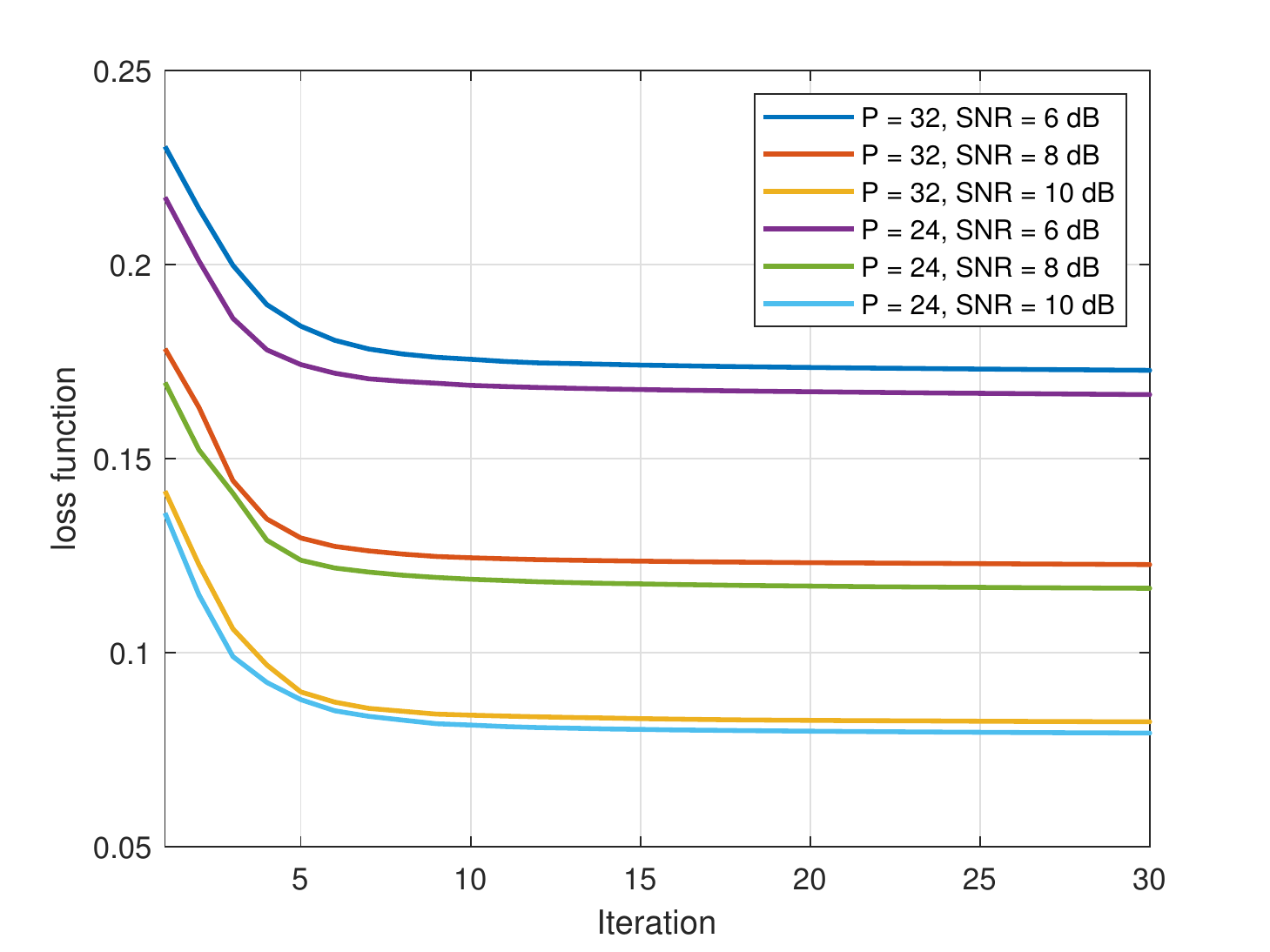}
	\vspace*{-1em}
	\caption{ Objective function with respect to the number of iterations.
	}
	\label{img:iteration}
	\vspace*{-1em}
\end{figure}
{	
The convergence behavior of the proposed P-SIGW algorithm is shown in Fig. \ref{img:iteration}. The objective function $\|\bar{\mb{Y}} - \tilde{\mb{\Psi}}(\hat{\bm{\theta}}, \hat{\mb{r}})\hat{\mb{G}} \|_F^2$ decreases monotonically over iteration. Under different parameters, the algorithm convergence is guaranteed. The simulation results are consistent with the convergence analysis in section \ref{sec:4-3}.  Taking into account the algorithm performance and complexity, we set the maximum iterations $N_{\text{iter}}$ as 10 in the following simulations.
}

{
	\begin{figure}
		\centering
		\includegraphics[width=3.5in]{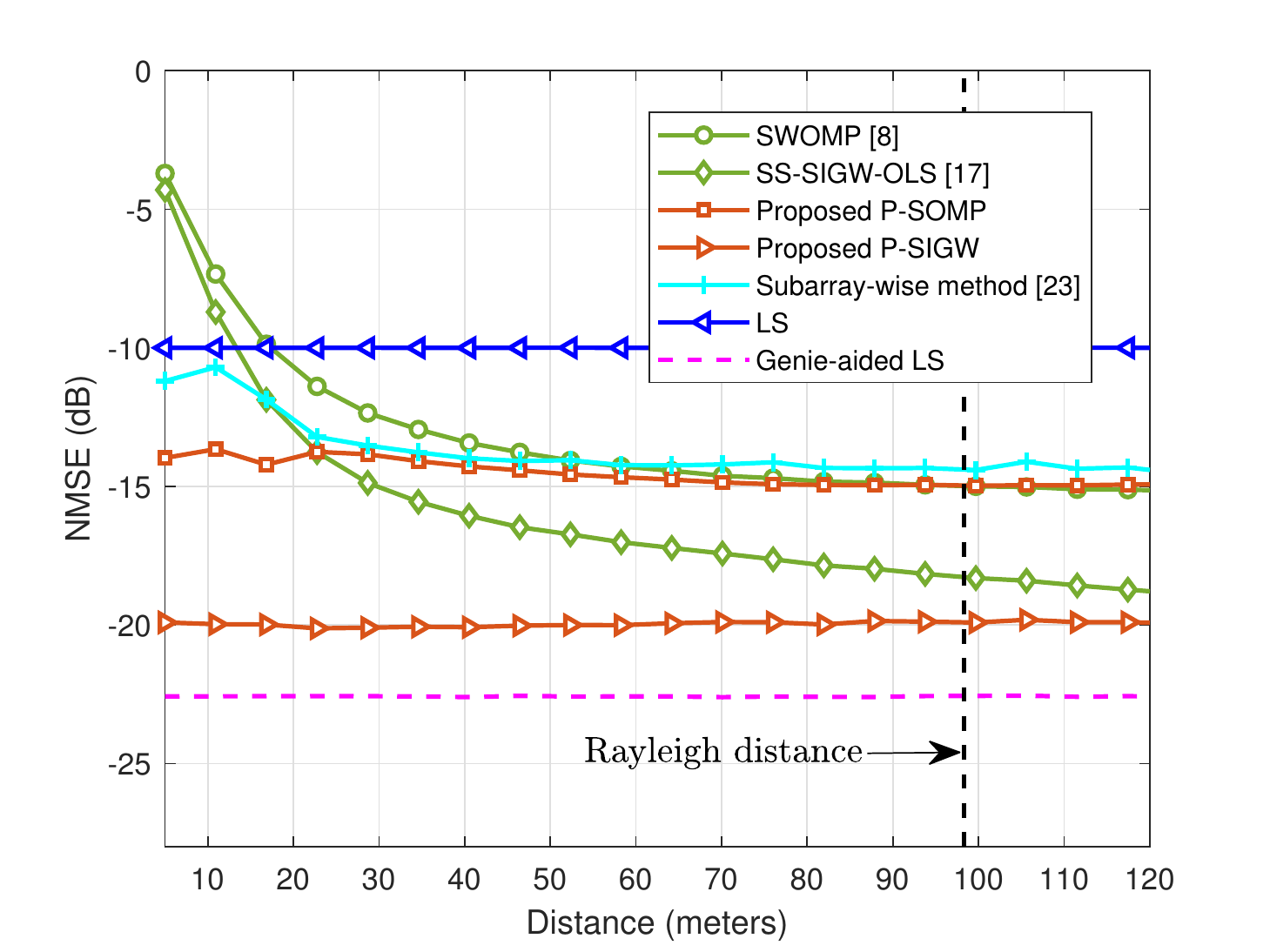}
		\vspace*{-1em}
		\caption{ NMSE performance comparison against the distance.
		}
		\label{img:nmse_dis}
		\vspace*{-1em}
	\end{figure}
The NMSE performance with respect to distance is evaluated in Fig. \ref{img:nmse_dis}. We compare 
the proposed on-grid polar-domain P-SOMP algorithm and the off-grid polar-domain P-SIGW algorithm with the existing methods, including the on-grid angular-domain SW-OMP algorithm \cite{SWOMP_Robert18}, the off-grid angular-domain SS-SIGW-OLS algorithm \cite{WideCT_Go21}, the subarray-wise near-field localization method proposed in \cite{NearCE_Han2020}, 
and the LS method. Moreover, the Genie-aid LS method, where the true distances and angles of the receivers and scatters are assumed to be available, is also compared as the NMSE performance bound.
The SNR is 10 dB, and the pilot length is $P = 32$, i.e., the compressive ratio is $\frac{PN_{\text {RF}}}{N} = \frac{1}{2}$.  {The distances between the BS and the users or scatters are increasing from 3 meters to 120 meters.} 	Since the number of antennas is $N = 256$, and the frequency is $f_c = 100$ GHz, then the Rayleigh distance is around 100 meters.
To highlight the impact of the near-field, we ignore the large-scale path loss in the channel. It can be observed from Fig. \ref{img:nmse_dis} that as the distance decreases, all of the far-field algorithms will suffer severe performance degradation, especially when the distance is lower than the Rayleigh distance. 
As for the subarray-wise near-field localization method, this method is equivalent to uniformly sampling multiple distances from the polar-domain, its NMSE performance is not robust to distance.
 By contrast, the proposed P-SOMP and P-SIGW algorithms outperform existing methods, and are robust to the small distance. Moreover, since the P-SIGW algorithm refines the estimated channel parameters with a much higher resolution, its NMSE performance is much better than the P-SOMP algorithm.

}

\begin{figure}
	\centering
	\subfigure[]
	{\includegraphics[width=3.5in]{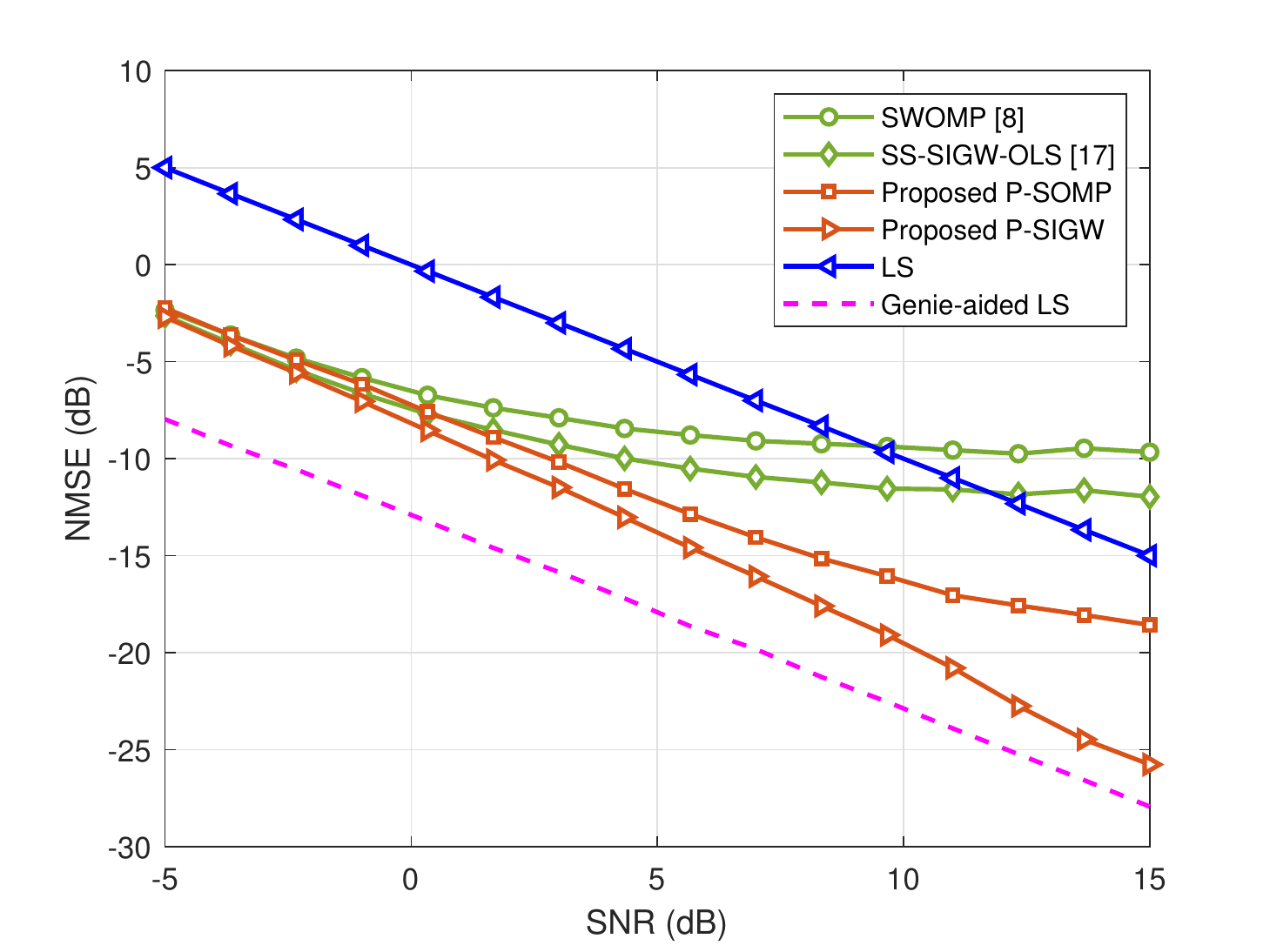}}
	\subfigure[]
	{\includegraphics[width=3.5in]{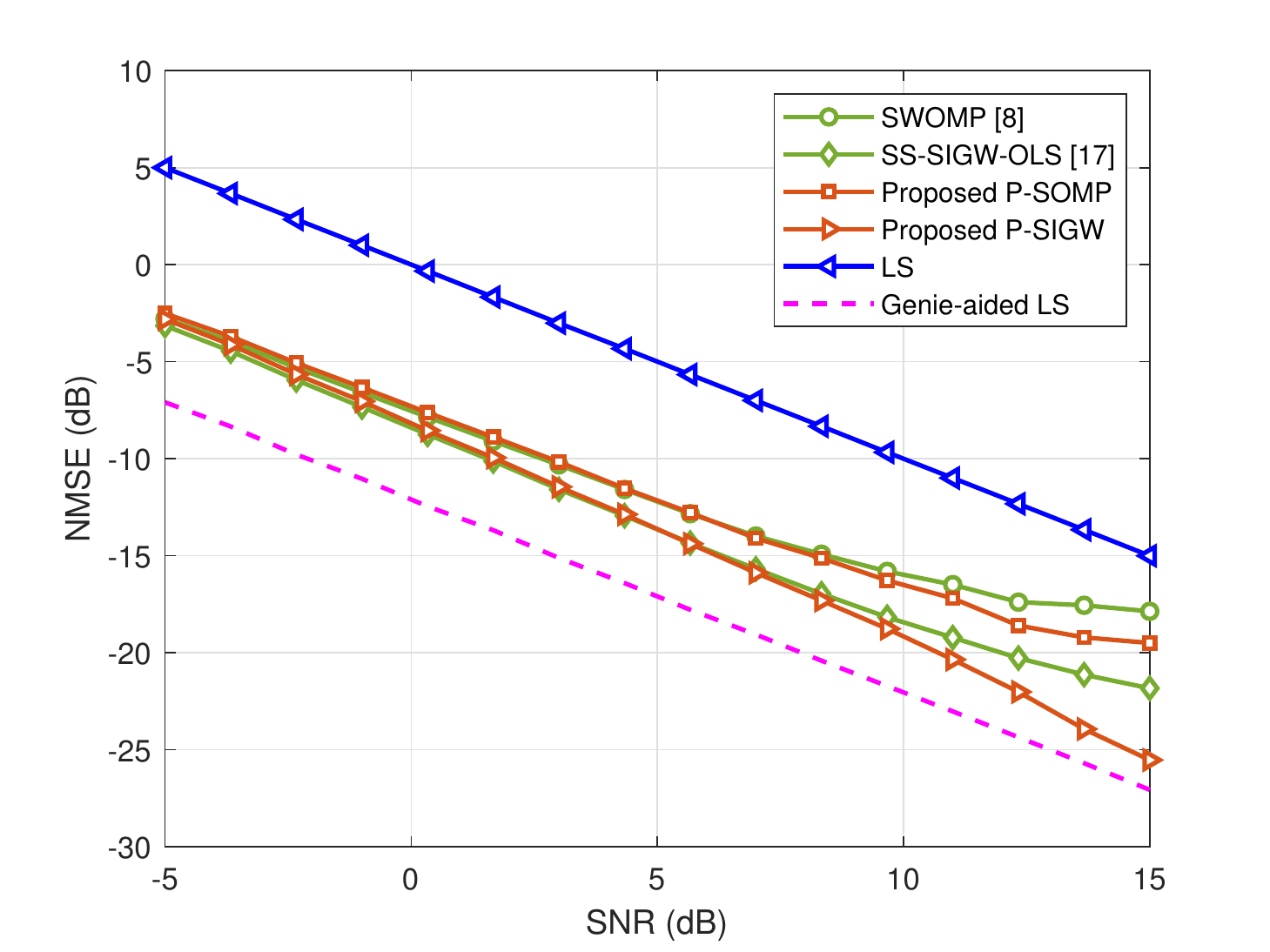}} 
	\\ 
	\centering
	\caption{The NMSE performance comparison against the SNR, where the distances between BS and the users or scatters are chosen from  $\mathcal{U}(5\:\text{m}, 10 \:\text{m})$ in (a), and the distances are chosen from  $\mathcal{U}(100\:\text{m}, 120 \:\text{m})$ in (b).} 
	\vspace*{-1em}
	\label{img:nmse_snr}
\end{figure}
	Then, Fig. \ref{img:nmse_snr} compares NMSE performance against SNR, where the length of pilot is $P = 32$. In Fig. \ref{img:nmse_snr} (a), the distances between the BS and the users or  scatters are randomly sampled from $\mathcal{U}(5\:\text{m}, 10 \:\text{m})$. In Fig. \ref{img:nmse_snr} (b), the corresponding distances are randomly sampled from  $\mathcal{U}(100 \:\text{m}, 120 \:\text{m})$. It can be observed from Fig. \ref{img:nmse_snr} (a) that when the distance is small, the proposed near-field channel estimation schemes significantly outperform existing far-field channel estimation algorithms at all considered SNR. 	
	For far-field scenario in Fig. \ref{img:nmse_snr} (b), the proposed near-field channel estimation schemes can achieve the similar NMSE performance compared with the existing angular-domain based algorithms. The reason is that, the designed polar-domain transform matrix $\mb{W}$ also samples distances in the far-field, e.g. $s = 0$ in (\ref{eq:distance}), so the polar-domain transform matrix can also extract the far-field information. Moreover, since the Fresnel approximation is more accurate than the far-field approximation, when the SNR is high ($>$ 10 dB), the proposed methods outperform the far-field methods, even for far-field channels, and approaches the performance bound achieved by Genie-aided LS scheme.
	In conclusion, the proposed near-field channel estimation algorithms can accurately recover the channel in both the near-field and far-field, while the existing far-field channel estimation algorithms can only accurately recover the far-field channel.
 
\begin{figure}
	\centering
	\subfigure[]
	{\includegraphics[width=3.5in]{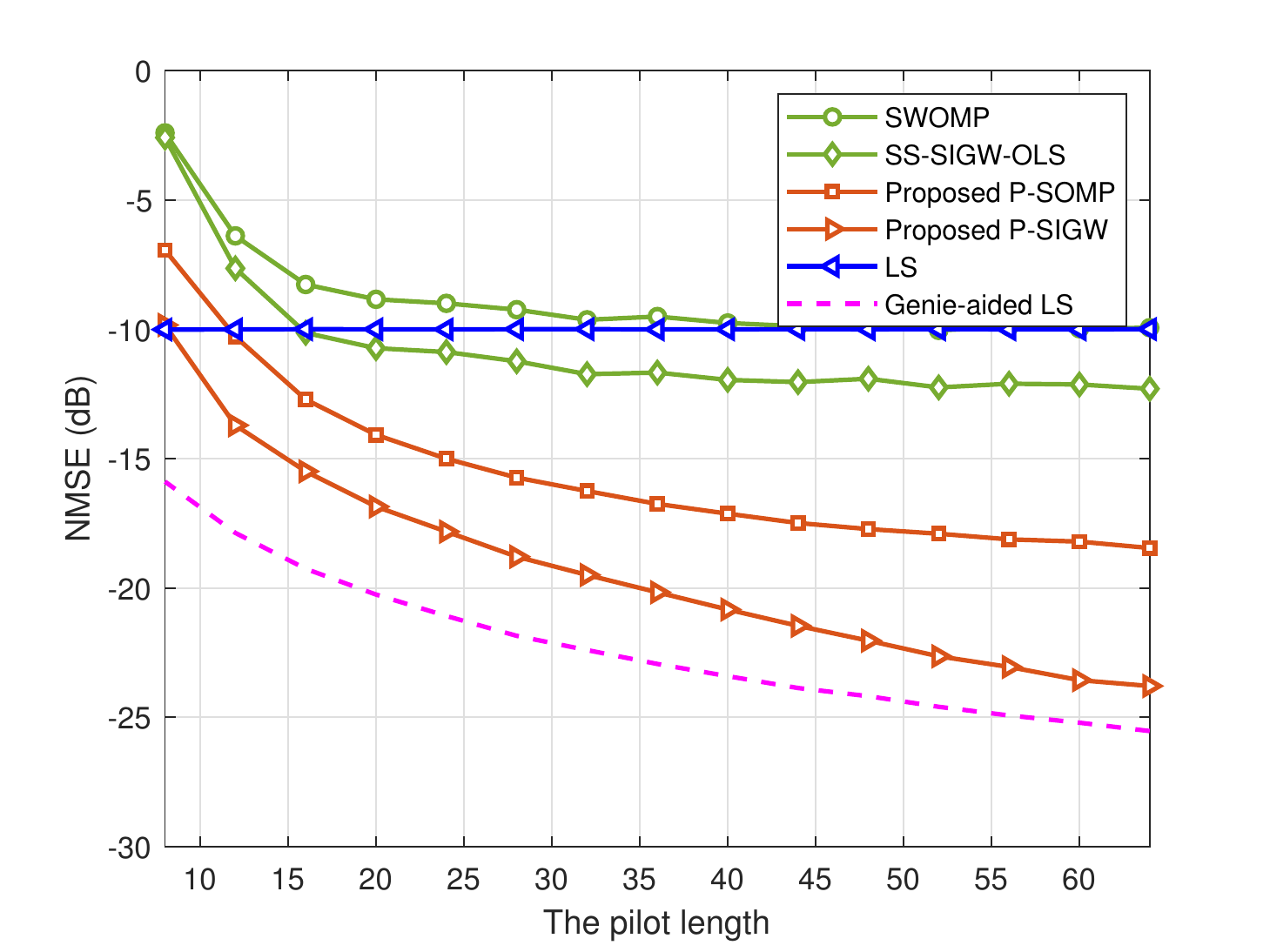}}
	\subfigure[]
	{\includegraphics[width=3.5in]{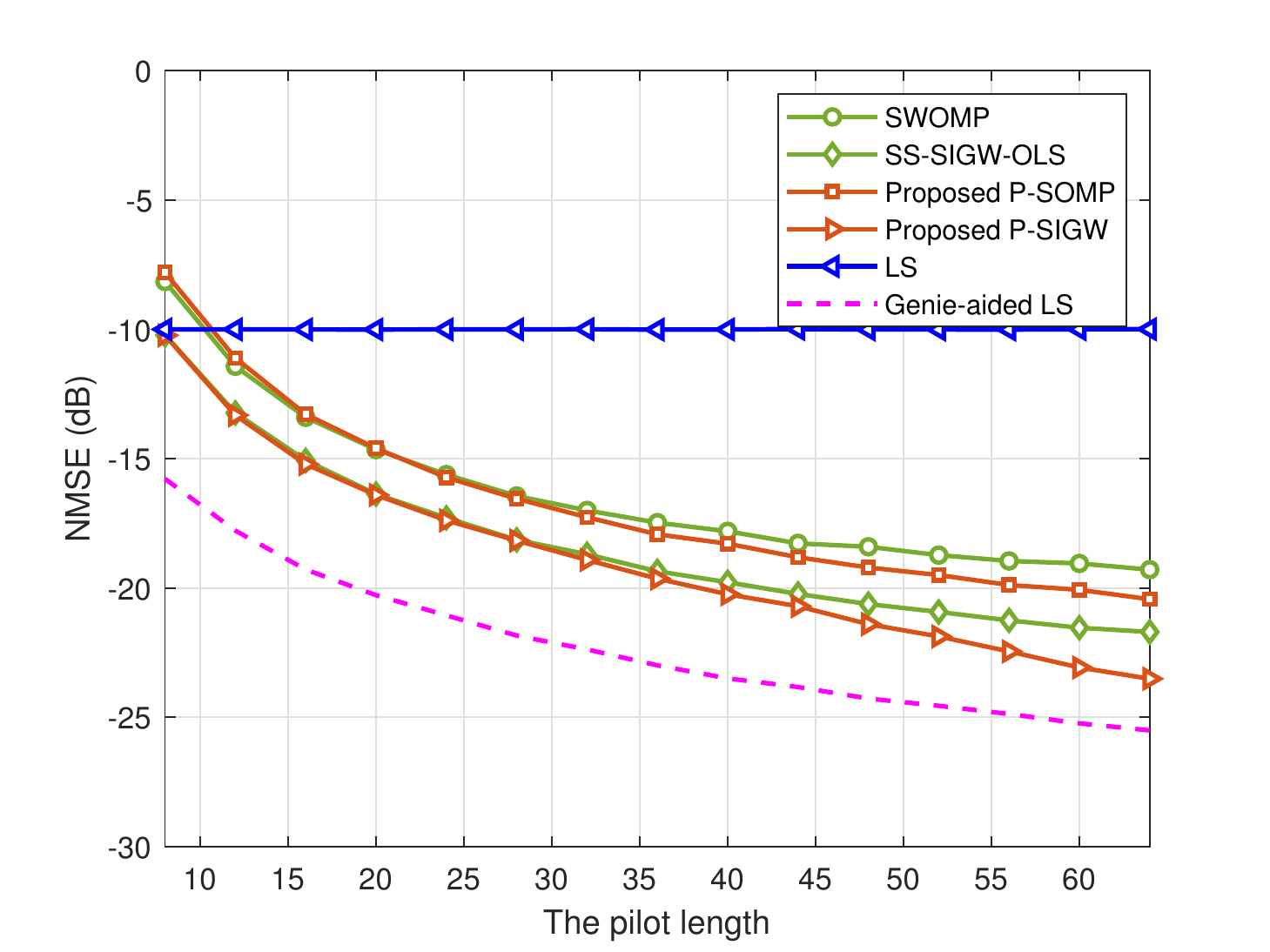}} 
	\\ 
	\centering
	\caption{The NMSE performance against the length of pilot, where the distances between the BS and the users or the scatters are chosen from  $\mathcal{U}(5\:\text{m}, 10 \:\text{m})$ in (a), and the distances are chosen from  $\mathcal{U}(100\:\text{m}, 120 \:\text{m})$ in (b).
	} 
\vspace*{-1em}
	\label{img:nmse_q}
\end{figure}

	{
	Fig. \ref{img:nmse_q} provides the NMSE performance against the pilot length $P$, where the SNR is 10 dB. In Fig. \ref{img:nmse_snr} (a), the distances are randomly sampled from $\mathcal{U}(5 \:\text{m}, 10 \:\text{m})$, while in Fig. \ref{img:nmse_snr} (b), the distances are randomly sampled from  $\mathcal{U}(100 \:\text{m}, 120 \:\text{m})$.
	 The length of pilot sequence $P$ is increasing from 8 to 64, so that the compressive ratio $\frac{PN_{\text {RF}}}{N}$ is increasing from $\frac{1}{8}$ to 1. It can be observed from Fig. \ref{img:nmse_q} that the NMSE performance achieved by all considered schemes improves as the pilot length $P$ becomes longer. As shown in Fig. \ref{img:nmse_q} (a), when the distance is small, the proposed P-SOMP and P-SIGW algorithms significantly outperform other angular-domain based algorithms. For example, when the pilot length is $P = 24$ or $P = 32$, the performance gap between the proposed P-SIGW algorithm and the existing algorithms is quite large. This indicates that the proposed algorithms can reduce the overhead for near-field XL-MIMO channel estimation. Moreover, for far-field scenario in Fig. \ref{img:nmse_q} (b), when the pilot length $P$ is shorter than 32, the NMSE achieved by the proposed near-field channel estimation schemes and existing far-field channel estimation schemes are similar, for both the on-grid and off-grid conditions. When the pilot length $P$ is longer than 32, the NMSE performance of the proposed algorithms slightly outperform the far-field algorithms. This is because the Fresnel approximation adopted in the polar-domain transform is more accurate than the far-field approximation.
	In conclusion, the proposed near-field channel estimation schemes can accurately recover the channel both in the near-field and far-field with low pilot overhead.
}

	\begin{figure}
		\centering
		\includegraphics[width=3.5in]{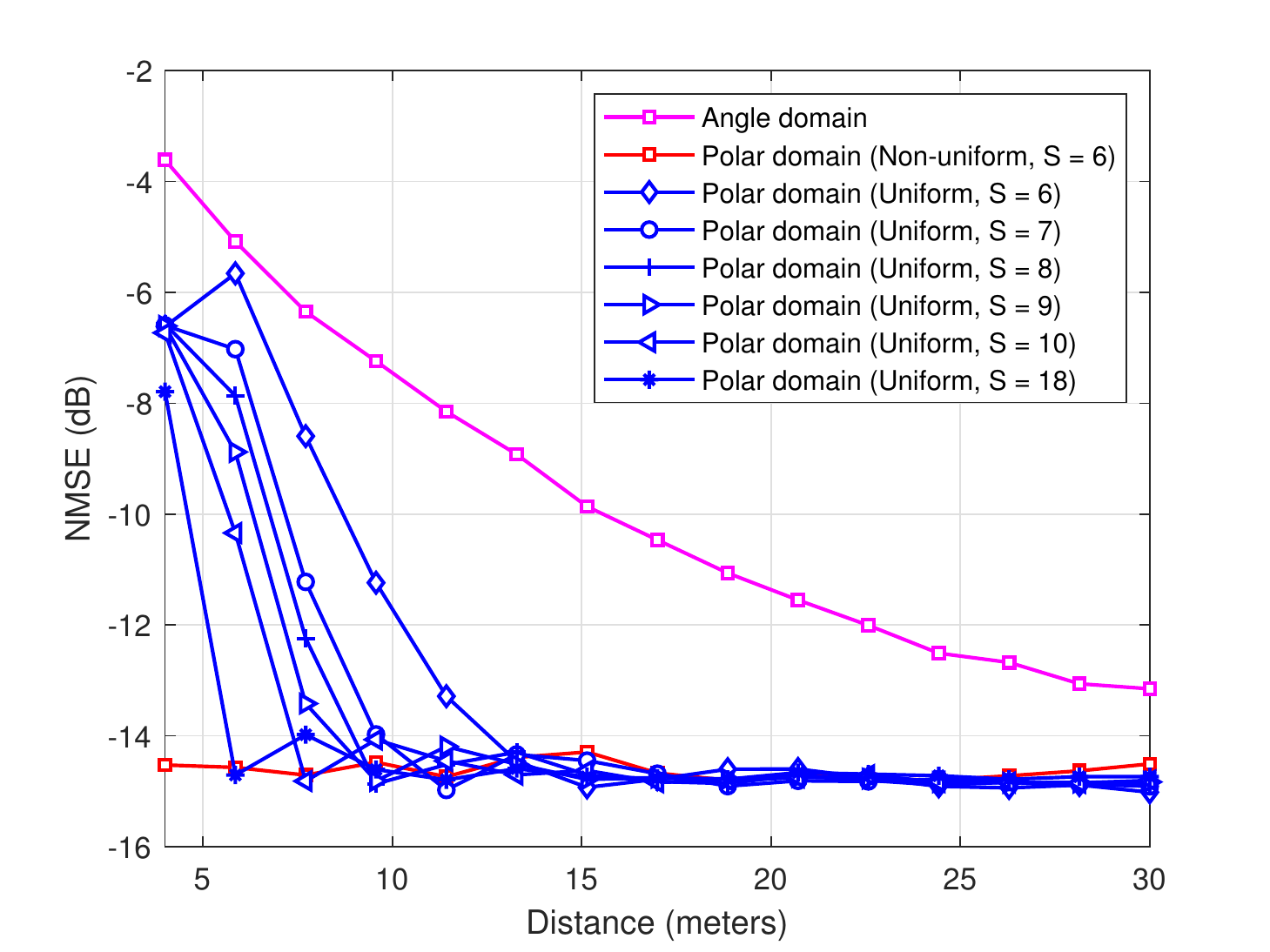}
		\vspace*{-1em}
		\caption{ NMSE performance comparison between the proposed non-uniform distance-sampling method and the uniform distance-sampling method.
		}
		\label{img:non_vs_uni}
		\vspace*{-1em}
	\end{figure}
{
	Finally, to verify the conclusion proved in (\ref{eq:distance}) that the distances should be sampled non-uniformly, we evaluate the NMSE performance achieved by the P-SOMP method with uniform distance-sampling method in Fig. \ref{img:non_vs_uni}. 
	For the uniform distance-sampling method, on each angle, $S$ distances are uniformly sampled from the predefined range $[\rho_{\text{min}}, \rho_{\text{max}}]$, i.e., $r_s = \rho_{\text{min}} + \frac{s}{S}(\rho_{\text{max}} - \rho_{\text{min}}), s=0,1,\cdots S-1 $, where we set $\rho_{\text{min}}$ as 3 meters and  $\rho_{\text{max}}$ as the Rayleigh distance. As shown in Fig. \ref{img:non_vs_uni}, although the estimation performance can be improved to some degree by uniformly sampling multiple distances in the polar domain,  the NMSE performance is severely degraded when the distance is small, even if the number of sampled distances $S$ on each angle is 18. However, with only $S = 6$ distance rings, the NMSE performance achieved by the proposed non-uniform distance-sampling method is robust for all considered distances.
}

	\section{Conclusions} \label{sec:6}
	In this paper, we have investigated the near-field channel estimation problem in XL-MIMO systems with hybrid precoding for the first time, where the near-field channel property was taken into account. 
	The energy spread effect for the near-field channel in the angular domain was revealed at first, where one near-field path component would spread towards multiple angles, and thus the angular-domain sparsity was not achievable in the near-field region. We further showed that the energy spread effect would severely degrade the performance of existing compressed sensing based channel estimation algorithms. 	
	To address this issue, we have proposed a polar-domain representation of the near-field XL-MIMO channel and designed the angular and distance sampling rules for the polar-domain transform matrix.
	Since the polar-domain transform matrix was able to simultaneously extract information of the angle and distance, both the far-field channel and near-field channel will be sparse in the polar domain.  
	Based on this polar-domain sparsity, an on-grid P-SOMP algorithm and an off-grid P-SIGW algorithm were proposed to estimate the near-field XL-MIMO channels.
	Simulation results show that in the near-field region, our proposed near-field channel estimation  schemes can achieve much better NMSE performance than existing far-field channel estimation schemes. In addition, the proposed near-field channel estimation schemes also performed well in the far-field region. 
	For future works, one may consider extending the proposed polar-domain representation to the near-field channel estimation in reconfigurable intelligent surface (RIS) aided communications \cite{RISCE_Wei21}. 
	
	\section*{Appendix A. Proof of Lemma 1}

	Substituting $k_c = \frac{2\pi}{\lambda_c}$ and $\theta_p = \theta_q$ into (\ref{eq:afp}), we get
	\begin{align}
	f(\theta, \theta, r_p, r_q) &= \left|\frac{1}{N}\sum_{n = -(N-1)/2}^{(N-1)/2}e^{j\pi n^2 \frac{d^2(1-\theta^2)}{\lambda_c}\left(\frac{1}{r_p} - \frac{1}{r_q}\right)} \right| \notag\\ &= \left|F(x)\right|,
	\end{align}
	where $x = \frac{d^2(1-\theta^2)}{\lambda_c}\left(\frac{1}{r_p} - \frac{1}{r_q}\right)$, and the function $F(x)$ is 
	\begin{align} \label{eq:F0}
	F(x) &= \frac{1}{N}\sum_{n = -(N-1)/2}^{(N-1)/2}e^{j\pi n^2 x} \approx \frac{1}{N}\int_{-N/2}^{N/2}e^{j\pi n^2 x} \text{d}n. 
	\end{align}
	If $r_p \le r_q$, then $x = \frac{d^2(1-\theta^2)}{\lambda_c}\left(\frac{1}{r_p} - \frac{1}{r_q}\right)$ is larger than 0, so we have
	\begin{align}
	F(x) & \overset{ (a) }{\approx} 
	\frac{2}{\sqrt{2x}N}\int_{0}^{\sqrt{2x}N/2}e^{j\frac{\pi}{2}t^2}\text{d}t 
 \notag  \\
	& = 
		\frac{\int_{0}^{\sqrt{2x}N/2}\cos(\frac{\pi}{2}t^2)\text{d}t +
			j\int_{0}^{\sqrt{2x}N/2}\sin(\frac{\pi}{2}t^2)\text{d}t
	}{\sqrt{2x}N/2}, \label{eq:F1}
	\end{align}
	where (a) is derived by letting $n^2x = \frac{1}{2}t^2$ in (\ref{eq:F0}). Note that $\int_{0}^{\sqrt{2x}N/2}\cos(\frac{\pi}{2}t^2)\text{d}t$ and $\int_{0}^{\sqrt{2x}N/2}\sin(\frac{\pi}{2}t^2)\text{d}t$ are Fresnel functions \cite{fresnel_Sherman1962}. We denote $\beta = \frac{\sqrt{2x}N}{2} $, $C(\beta) = \int_{0}^{\beta}\cos(\frac{\pi}{2}t^2)\text{d}t$, and $S(\beta) = \int_{0}^{\beta}\sin(\frac{\pi}{2}t^2)\text{d}t$, then the function $F(x)$ can rewritten as 
	\begin{align}
	F(x) =  \frac{C(\beta) + jS(\beta)}{\beta} 
	= G(\beta),  \label{eq:F}
	\end{align}
		where $\beta = \frac{\sqrt{2x}N}{2} = \sqrt{\frac{N^2d^2(1 - \theta^2)}{2\lambda_c} \left( \frac{1}{r_p} - \frac{1}{r_q}\right)}$.
	
	Furthermore, if $r_p > r_q$, then we have $ -x = \frac{d^2(1-\theta^2)}{\lambda_c}\left(\frac{1}{r_q} - \frac{1}{r_p}\right) > 0$. Similar to the derivation of (\ref{eq:F}), we can obtain that if $-x > 0$, then 
	$F(x) = G^*(\beta)$, where $\beta = \sqrt{\frac{N^2d^2(1 - \theta^2)}{2\lambda_c} \left( \frac{1}{r_q} - \frac{1}{r_p}\right)}$.
	 
	To sum up, the column coherence can be approximated as $f(\theta, \theta, r_p, r_q) \approx \left|F(x)\right| \approx \left|G(\beta)\right|$, where $\beta = \sqrt{\frac{N^2d^2(1 - \theta^2)}{2\lambda_c} \left| \frac{1}{r_q} - \frac{1}{r_p}\right|}$. Therefore, the proof of \textbf{Lemma 1} is completed.

\section*{Appendix B. Derivation Of The Gradient}
In this appendix, the explicit derivation of the gradient of $ \mathcal{L}(\hat{\bm{\theta}}, \hat{\mb{r}})$ 	with respect to $\hat{\mb{r}} = [\hat{r}_1, \hat{r}_2, \cdots, \hat{r}_{\hat L}]$ and $\hat{\bm{\theta}} = [\hat{\theta}_1, \hat{\theta}_2, \cdots, \hat{\theta}_{\hat L}]$ is provided. For the $l$-th angle $\hat{\theta}_l$, the gradient of $ \mathcal{L}(\hat{\bm{\theta}}, \hat{\mb{r}})$ is given by
\begin{align} \label{eq:ap2_1}
\frac{\partial \mathcal{L}(\hat{\bm{\theta}}, \hat{\mb{r}})}{\partial \hat{\theta}_l} = -\text{Tr}\left\{\bar{\mb{Y}}^H\frac{\partial \mb{P}(\hat{\bm{\theta}}, \hat{\mb{r}})}{\partial \hat{\theta}_l}\bar{\mb{Y}}\right\}.
\end{align}
For expression simplicity, in the following derivation, we ignore the term $(\hat{\bm{\theta}}, \hat{\mb{r}})$. Since $\mb{P} = \tilde{\mb{\Psi}}\tilde{\mb{\Psi}}^{\dagger}$ and $\tilde{\mb{\Psi}}^{\dagger} = \left( \tilde{\mb{\Psi}}^H\tilde{\mb{\Psi}} \right)^{-1}\tilde{\mb{\Psi}}^H$, the gradient of $\mb{P}$ is given by
\begin{align} \label{eq:ap2_4}
\frac{\partial \mb{P}}{\partial \hat{\theta}_l} = \frac{\partial \tilde{\mb{\Psi}}}{\partial \hat{\theta}_l} \left( \tilde{\mb{\Psi}}^H\tilde{\mb{\Psi}} \right)^{-1}\tilde{\mb{\Psi}}^H &+ \tilde{\mb{\Psi}}\frac{\partial \left( \tilde{\mb{\Psi}}^H\tilde{\mb{\Psi}} \right)^{-1}}{\partial \hat{\theta}_l}\tilde{\mb{\Psi}}^H \notag\\ &+ \tilde{\mb{\Psi}}\left(\tilde{\mb{\Psi}}^H\tilde{\mb{\Psi}} \right)^{-1}\frac{\partial \tilde{\mb{\Psi}}^H}{\partial \hat{\theta}_l}.
\end{align}
According to the inverse matrix differentiation law that $d\mb{A}^{-1} = -\mb{A}^{-1}(d\mb{A})\mb{A}^{-1}$,
the gradient of $\left( \tilde{\mb{\Psi}}^H\tilde{\mb{\Psi}} \right)^{-1}$ is given by
\begin{align} \label{eq:ap2_5}
&\frac{\partial \left( \tilde{\mb{\Psi}}^H\tilde{\mb{\Psi}} \right)^{-1}}{\partial \hat{\theta}_l}= \notag\\&- \left( \tilde{\mb{\Psi}}^H\tilde{\mb{\Psi}} \right)^{-1}
\left(
\frac{\partial  \tilde{\mb{\Psi}}^H}{\partial \hat{\theta}_l}\tilde{\mb{\Psi}}  + \tilde{\mb{\Psi}}^H\frac{\partial  \tilde{\mb{\Psi}}}{\partial \hat{\theta}_l} 
\right)
\left( \tilde{\mb{\Psi}}^H\tilde{\mb{\Psi}} \right)^{-1}.
\end{align}
Since $\tilde{\mb{\Psi}}(\hat{\bm{\theta}}, \hat{\mb{r}}) = \mb{D}^{-1}\mb{A}\tilde{\mb{W}}(\hat{\bm{\theta}}, \hat{\mb{r}})$, the gradient of $\tilde{\mb{\Psi}}$ is given by
\begin{align} \label{eq:ap2_6}
\frac{\partial  \tilde{\mb{\Psi}}}{\partial \hat{\theta}_l} = \mb{D}^{-1}\mb{A}\frac{\partial  \tilde{\mb{W}}}{\partial \hat{\theta}_l},
\end{align}
where the gradient of $\tilde{\mb{W}}$ is
\begin{align}\label{eq:ap2_2}
\frac{\partial  \tilde{\mb{W}}}{\partial \hat{\theta}_l} = \left[\mb{0}, \cdots, \mb{0}, 
\frac{\partial  \mb{b}(\hat{\theta}_l, \hat{r}_l)}{\partial \hat{\theta}_l},\mb{0},  \cdots,\mb{0} \right].
\end{align}
Combining (\ref{eq:ap2_1})-(\ref{eq:ap2_2}), we can derive the gradient of $ \mathcal{L}(\hat{\bm{\theta}}, \hat{\mb{r}})$ against the $\hat{\theta}_l$.  The same
procedure can be carried out to calculate the gradients of the remaining angle parameters and distance parameters. The only difference for the gradients of $ \frac{1}{\hat{\mb{r}}}$ is that (\ref{eq:ap2_2}) is changed to

 \begin{align}\label{eq:ap2_3}
 \frac{\partial  \tilde{\mb{W}}}{\partial \frac{1}{\hat{r}_l}} = \left[\mb{0}, \cdots, \mb{0}, 
 \frac{\partial  \mb{b}(\hat{\theta}_l, \hat{r}_l)}{\partial \frac{1}{\hat{r}_l}},\mb{0},  \cdots,\mb{0} \right].
 \end{align}
Finally, stacking all of the distance and angle terms in a column vector, we can obtain the gradient $\nabla_{\frac{1}{\hat{\mb{r}}}}\mathcal{L}(\hat{\bm{\theta}}, \hat{\mb{r}})$ and $\nabla_{\hat{\bm{\theta}}}\mathcal{L}(\hat{\bm{\theta}}, \hat{\mb{r}})$.

	\bibliographystyle{IEEEtran}
	\bibliography{IEEEabrv,refs}

\begin{thebibliography}{10}
\providecommand{\url}[1]{#1}
\csname url@samestyle\endcsname
\providecommand{\newblock}{\relax}
\providecommand{\bibinfo}[2]{#2}
\providecommand{\BIBentrySTDinterwordspacing}{\spaceskip=0pt\relax}
\providecommand{\BIBentryALTinterwordstretchfactor}{4}
\providecommand{\BIBentryALTinterwordspacing}{\spaceskip=\fontdimen2\font plus
\BIBentryALTinterwordstretchfactor\fontdimen3\font minus
  \fontdimen4\font\relax}
\providecommand{\BIBforeignlanguage}[2]{{%
\expandafter\ifx\csname l@#1\endcsname\relax
\typeout{** WARNING: IEEEtran.bst: No hyphenation pattern has been}%
\typeout{** loaded for the language `#1'. Using the pattern for}%
\typeout{** the default language instead.}%
\else
\language=\csname l@#1\endcsname
\fi
#2}}
\providecommand{\BIBdecl}{\relax}
\BIBdecl

\bibitem{Cui2112:Near}
M.~Cui and L.~Dai, ``{Near-Field} channel estimation for extremely large-scale
  {MIMO} systems with hybrid precoding,'' in \emph{Proc. 2021 IEEE Global
  Communications Conference (IEEE GLOBECOM'21)}, Dec. 2021, pp. 1--6.

\bibitem{5Gwork_Par13}
T.~S. Rappaport, S.~Sun, R.~Mayzus, H.~Zhao, Y.~Azar, K.~Wang, G.~N. Wong,
  J.~K. Schulz, M.~Samimi, and F.~Gutierrez, ``Millimeter wave mobile
  communications for {5G} cellular: It will work!'' \emph{IEEE Access}, vol.~1,
  pp. 335--349, May 2013.

\bibitem{MIMO2_Hu2018}
S.~{Hu}, F.~{Rusek}, and O.~{Edfors}, ``Beyond massive {MIMO}: The potential of
  data transmission with large intelligent surfaces,'' \emph{IEEE Trans. Signal
  Process.}, vol.~66, no.~10, pp. 2746--2758, May 2018.

\bibitem{6Gchallenge_Rappaport2019}
T.~S. {Rappaport}, Y.~{Xing}, O.~{Kanhere}, S.~{Ju}, A.~{Madanayake},
  S.~{Mandal}, A.~{Alkhateeb}, and G.~C. {Trichopoulos}, ``Wireless
  communications and applications above 100 {GHz}: Opportunities and challenges
  for 6{G} and beyond,'' \emph{IEEE Access}, vol.~7, pp. 78\,729--78\,757,
  2019.

\bibitem{THzsurvey_Elayan2020}
H.~{Elayan}, O.~{Amin}, B.~{Shihada}, R.~M. {Shubair}, and M.~{Alouini},
  ``Terahertz band: The last piece of {RF} spectrum puzzle for communication
  systems,'' \emph{IEEE Open J. Commun. Society}, vol.~1, pp. 1--32, 2020.

\bibitem{NomaSWIPT_Dai19}
L.~Dai, B.~Wang, M.~Peng, and S.~Chen, ``Hybrid precoding-based millimeter-wave
  massive {MIMO-NOMA} with simultaneous wireless information and power
  transfer,'' \emph{IEEE J. Sel. Areas Commun.}, vol.~37, no.~1, pp. 131--141,
  Jan. 2019.

\bibitem{UMIMO_Han21}
B.~Ning, Z.~Tian, Z.~Chen, C.~Han, J.~Yuan, and S.~Li, ``Prospective
  beamforming technologies for ultra-massive {MIMO} in terahertz
  communications: A tutorial,'' \emph{arXiv preprint arXiv:2107.03032}, Jul.
  2021.

\bibitem{SWOMP_Robert18}
J.~Rodríguez-Fernández, N.~González-Prelcic, K.~Venugopal, and R.~W. Heath,
  ``Frequency-domain compressive channel estimation for frequency-selective
  hybrid millimeter wave {MIMO} systems,'' \emph{IEEE Trans. Wireless Commun.},
  vol.~17, no.~5, pp. 2946--2960, May 2018.

\bibitem{CE_OMP_Lee16}
J.~Lee, G.-T. Gil, and Y.~H. Lee, ``Channel estimation via orthogonal matching
  pursuit for hybrid {MIMO} systems in millimeter wave communications,''
  \emph{IEEE Trans. Commun.}, vol.~64, no.~6, pp. 2370--2386, Jun. 2016.

\bibitem{CE_SOMP_Gao16}
Z.~Gao, C.~Hu, L.~Dai, and Z.~Wang, ``Channel estimation for millimeter-wave
  massive {MIMO} with hybrid precoding over frequency-selective fading
  channels,'' \emph{IEEE Commun. Lett.}, vol.~20, no.~6, pp. 1259--1262, Jun.
  2016.

\bibitem{CE_SMP_Huang19}
C.~Huang, L.~Liu, C.~Yuen, and S.~Sun, ``Iterative channel estimation using
  {LSE} and sparse message passing for mm{Wave} {MIMO} systems,'' \emph{IEEE
  Trans. Signal Process.}, vol.~67, no.~1, pp. 245--259, Jan. 2019.

\bibitem{LAMP_Wei21}
X.~Wei, C.~Hu, and L.~Dai, ``Deep learning for beamspace channel estimation in
  millimeter-wave massive {MIMO} systems,'' \emph{IEEE Trans. Commun.},
  vol.~69, no.~1, pp. 182--193, Jan. 2021.

\bibitem{WidebeamCE_Gao2019}
X.~{Gao}, L.~{Dai}, S.~{Zhou}, A.~M. {Sayeed}, and L.~{Hanzo}, ``Wideband
  beamspace channel estimation for millimeter-wave {MIMO} systems relying on
  lens antenna arrays,'' \emph{IEEE Trans. Signal Process.}, vol.~67, no.~18,
  pp. 4809--4824, Sep. 2019.

\bibitem{UGSWOMP_Ro17}
J.~Rodríguez-Fernández, N.~González-Prelcic, and R.~W. Heath, ``A
  compressive sensing-maximum likelihood approach for off-grid wideband channel
  estimation at {mmWave},'' in \emph{Proc. 2017 IEEE 7th International Workshop
  on Computational Advances in Multi-Sensor Adaptive Processing (IEEE CAMSAP)},
  Dec. 2017, pp. 1--5.

\bibitem{TD_Zhou17}
Z.~Zhou, J.~Fang, L.~Yang, H.~Li, Z.~Chen, and R.~S. Blum, ``Low-rank tensor
  decomposition-aided channel estimation for millimeter wave {MIMO-OFDM}
  systems,'' \emph{IEEE J. Sel. Areas Commun.}, vol.~35, no.~7, pp. 1524--1538,
  Jul. 2017.

\bibitem{CESR_Hu18}
C.~Hu, L.~Dai, T.~Mir, Z.~Gao, and J.~Fang, ``Super-resolution channel
  estimation for {mmWave} massive {MIMO} with hybrid precoding,'' \emph{IEEE
  Trans. Veh. Technol.}, vol.~67, no.~9, pp. 8954--8958, Sep. 2018.

\bibitem{WideCT_Go21}
N.~González-Prelcic, H.~Xie, J.~Palacios, and T.~Shimizu, ``Wideband channel
  tracking and hybrid precoding for {mmWave} {MIMO} systems,'' \emph{IEEE
  Trans. Wireless Commun.}, vol.~20, no.~4, pp. 2161--2174, Apr. 2021.

\bibitem{fresnel_Selvan2017}
K.~T. {Selvan} and R.~{Janaswamy}, ``Fraunhofer and fresnel distances: Unified
  derivation for aperture antennas,'' \emph{IEEE Antennas Propag. Mag.},
  vol.~59, no.~4, pp. 12--15, Aug. 2017.

\bibitem{NearLoS_Zhou2015}
Z.~{Zhou}, X.~{Gao}, J.~{Fang}, and Z.~{Chen}, ``Spherical wave channel and
  analysis for large linear array in {LoS} conditions,'' in \emph{Proc. IEEE
  Globecom Workshops 2015}, Dec. 2015, pp. 1--6.

\bibitem{SPlocalization_Wei18}
W.~Zuo, J.~Xin, N.~Zheng, and A.~Sano, ``Subspace-based localization of
  far-field and near-field signals without eigendecomposition,'' \emph{IEEE
  Trans. signal process.}, vol.~66, no.~17, pp. 4461--4476, Sep. 2018.

\bibitem{NearCE_Liu2020}
H.~Liu, H.~Meng, L.~Gan, D.~Li, Y.~Zhou, and T.-K. Truong, ``Subspace and
  sparse reconstruction based near-field sources localization in uniform linear
  array,'' \emph{Digital Signal Process.}, vol. 106, p. 102824, 2020.

\bibitem{NearCE_Fried2019}
B.~{Friedlander}, ``Localization of signals in the near-field of an antenna
  array,'' \emph{IEEE Trans. Signal Process.}, vol.~67, no.~15, pp. 3885--3893,
  Aug. 2019.

\bibitem{NearCE_Han2020}
Y.~{Han}, S.~{Jin}, C.~{Wen}, and X.~{Ma}, ``Channel estimation for extremely
  large-scale massive {MIMO} systems,'' \emph{IEEE Wireless Commun. Lett.},
  vol.~9, no.~5, pp. 633--637, May 2020.

\bibitem{FundWC_Tse2015}
D.~{Tse} and P.~{Viswanath}, \emph{Fundamentals of Wireless
  Communication}.\hskip 1em plus 0.5em minus 0.4em\relax Cambridge, U.K.:
  Cambridge Univ. Press, 2005.

\bibitem{CE_Wang21}
H.~Wang, J.~Fang, P.~Wang, G.~Yue, and H.~Li, ``Efficient beamforming training
  and channel estimation for millimeter wave {OFDM} systems,'' \emph{IEEE
  Trans. Wireless Commun.}, vol.~20, no.~5, pp. 2805--2819, May 2021.

\bibitem{fresnel_Sherman1962}
J.~{Sherman}, ``Properties of focused apertures in the {Fresnel} region,''
  \emph{IRE Trans. Antennas Propag.}, vol.~10, no.~4, pp. 399--408, Jul. 1962.

\bibitem{CS_Ba10}
W.~U. Bajwa, J.~Haupt, A.~M. Sayeed, and R.~Nowak, ``Compressed channel
  sensing: A new approach to estimating sparse multipath channels,''
  \emph{Proc. IEEE}, vol.~98, no.~6, pp. 1058--1076, Jun. 2010.

\bibitem{AO_Bezdek2002}
J.~C. Bezdek and R.~J. Hathaway, ``Some notes on alternating optimization,'' in
  \emph{Advances in Soft Computing}, Feb. 2002, pp. 288--300.

\bibitem{AO_Bezdek2003}
D.~B. Fogel and C.~J. Robinson, ``Two new convergence results for alternating
  optimization,'' in \emph{Computational Intelligence: The Experts Speak},
  2003, pp. 149--164.

\bibitem{RISCE_Wei21}
L.~Wei, C.~Huang, G.~C. Alexandropoulos, C.~Yuen, Z.~Zhang, and M.~Debbah,
  ``Channel estimation for {RIS}-empowered multi-user {MISO} wireless
  communications,'' \emph{IEEE Trans. Commun.}, vol.~69, no.~6, pp. 4144--4157,
  Jun. 2021.

\end{thebibliography}

\end{document}